\documentclass[longauth]{aa}
\usepackage{graphicx}
\usepackage{txfonts}
\usepackage[dvipsnames]{xcolor}
\usepackage{orcidlink}
\usepackage{colortbl}
\usepackage[rightcaption]{sidecap}
\sidecaptionvpos{figure}{c}

\usepackage{hyperref}  
\hypersetup{colorlinks=true,linkcolor=[rgb]{1.,0.2,0.2},citecolor=[rgb]{0.1,0.4,1.},filecolor=[rgb]{0.7,0.2,0.2},urlcolor=[rgb]{0.7,0.2,0.2}}

\usepackage{color}
\definecolor{blue}{rgb}{0., 0., 1}

\def\cgs{\rm{erg}\ \rm{cm^{-2}}\ \rm{s^{-1}}}
\def\ergs{\rm{erg}\ \rm{s^{-1}}}

\def\Msun{M$_\odot$}

\def\Hei{He\,{\sc i}}

\def\Oiii{[O\,{\sc iii}]}

\def\Oip{O\,{\sc i}}
\def\Oi{[O\,{\sc i}]}

\def\Ha{H$\alpha$}
\def\Hb{H$\beta$}

\def\Pab{Pa$\beta$}
\def\Pag{Pa$\gamma$}

\def\kms{km\,s$^{-1}$}

\def\lsim{\mathrel{\rlap{\lower 3pt \hbox{$\sim$}} \raise 2.0pt \hbox{$<$}}}
\def\gsim{\mathrel{\rlap{\lower 3pt \hbox{$\sim$}} \raise 2.0pt \hbox{$>$}}}

\begin{document}

\authorrunning{Loiacono F. et al.}
\titlerunning{A big red dot at cosmic noon}
\title{A big red dot at cosmic noon}
\author{
Federica Loiacono\inst{1}\thanks{ \email{federica.loiacono1@inaf.it}}$^{\orcidlink{0000-0002-8858-6784}}$, 
Roberto Gilli\inst{1}$^{\orcidlink{0000-0001-8121-6177}}$, 
Marco Mignoli\inst{1}$^{\orcidlink{0000-0002-9087-2835}}$, 
Giovanni Mazzolari\inst{17,1}$^{\orcidlink{0009-0005-7383-6655}}$, Roberto Decarli\inst{1}$^{\orcidlink{0000-0002-2662-8803}}$, Marcella Brusa\inst{2,1}$^{\orcidlink{0000-0002-5059-6848}}$, Francesco Calura\inst{1}$^{\orcidlink{0000-0002-6175-0871}}$, Marco Chiaberge\inst{3,4}$^{\orcidlink{0000-0003-1564-3802}}$, Andrea Comastri\inst{1}$^{\orcidlink{0000-0003-3451-9970}}$, Quirino D'Amato\inst{5}$^{\orcidlink{0000-0002-9948-0897}}$, Kazushi Iwasawa\inst{6,7}$^{\orcidlink{0000-0002-4923-3281}}$, Ignas Juodžbalis\inst{13,14}$^{\orcidlink{0009-0003-7423-8660}}$, Giorgio Lanzuisi\inst{1}$^{\orcidlink{0000-0001-9094-0984}}$, Roberto Maiolino\inst{13,14,15}$^{\orcidlink{0000-0002-4985-3819}}$, Stefano Marchesi\inst{2,8,1}$^{\orcidlink{0000-0001-5544-0749}}$, Colin Norman\inst{4,9}$^{\orcidlink{0000-0002-5222-5717}}$, Alessandro Peca\inst{10,11}$^{\orcidlink{0000-0003-2196-3298}}$, Isabella Prandoni\inst{16}$^{\orcidlink{0000-0001-9680-7092}}$, Matteo Sapori\inst{2}$^{\orcidlink{0009-0008-3970-4765}}$, Matilde Signorini\inst{5,12}$^{\orcidlink{0000-0002-8177-6905}}$, Paolo Tozzi\inst{5}$^{\orcidlink{0000-0003-3096-9966}}$, Eros Vanzella\inst{1}$^{\orcidlink{0000-0002-5057-135X}}$, Cristian Vignali\inst{2,1}$^{\orcidlink{0000-0002-8853-9611}}$, Fabio Vito\inst{1}$^{\orcidlink{0000-0003-0680-9305}}$, and Gianni Zamorani\inst{1}$^{\orcidlink{0000-0002-2318-301X}}$}
\institute{
INAF -- Osservatorio di Astrofisica e Scienza dello Spazio di Bologna, via Gobetti 93/3, I-40129, Bologna, Italy \and Dipartimento di Fisica e Astronomia, Università di Bologna, Via
Gobetti 93/2, I-40129 Bologna, Italy \and Space Telescope Science Institute for the European Space Agency (ESA), ESA Office, 3700 San Martin Drive, Baltimore, MD 21218,
USA \and The William H. Miller III Department of Physics and Astronomy, Johns Hopkins University, Baltimore, MD 21218, USA \and INAF – Arcetri Astrophysical Observatory, Largo E. Fermi 5, I50125, Florence, Italy \and Institut de Ciències del Cosmos (ICCUB), Universitat de Barcelona (IEEC-UB), Martí i Franquès, 1, 08028 Barcelona, Spain \and ICREA, Pg. Lluís Companys 23, 08010 Barcelona, Spain \and Department of Physics and Astronomy, Clemson University, Kinard Lab of Physics, Clemson, SC 29634-0978, USA \and Space Telescope Science Institute, 3700 San Martin Drive, Baltimore, MD, USA \and Eureka Scientific, 2452 Delmer Street, Suite 100, Oakland, CA 94602-3017, USA \and Department of Physics, Yale University, P.O. Box 208120, New Haven, CT 06520, USA \and Dipartimento di Matematica e Fisica, Univeristà di Roma 3, Via della Vasca Navale, 84, 00146 Roma, Italy \and Kavli Institute for Cosmology, University of Cambridge, Madingley Road, Cambridge CB3 OHA, UK \and Cavendish Laboratory – Astrophysics Group, University of Cambridge, 19 JJ Thomson Avenue, Cambridge CB3 OHE, UK \and Department of Physics and Astronomy, University College London, Gower Street, London WC1E 6BT, UK \and INAF - Istituto di Radioastronomia, Via Gobetti 101, I-40129 Bologna, Italy \and Max-Planck-Institut für extraterrestrische Physik (MPE), Gießenbachstraße 1, 85748 Garching, Germany}

\date{}
\abstract{We report the discovery of a little red dot (LRD), dubbed BiRD ('big red dot'), at $z \sim 2.33$ in the field around the $z\sim 6.3$ quasar SDSS J1030+0524. Using JWST/NIRCam images, we identified it as a bright outlier in the $F200W - F356W$ color versus $F356W$ magnitude diagram of point sources in the field. The NIRCam/WFSS spectrum reveals the emission from \Hei\ $\lambda\ 10830$ and \Pag\ line, both showing a narrow and a broad ($FWHM \gtrsim 2000\ \rm km\ s^{-1}$) component. The \Hei\ line is affected by an absorption feature, tracing dense gas with \Hei\ column density in the $2^3S$ level $N(\rm He\ I) \sim 0.5 - 1.2  \times 10^{14}\ \rm cm^{-2}$, depending on the location of the absorber, which is outflowing at the speed of $\Delta v =-830_{-148}^{+131}$ \kms. As observed in the majority of LRDs, BiRD does not show X-ray or radio emission down to $3.7 \times 10^{42}\ \rm erg\ s^{-1}$ and $3 \times 10^{39}\ \rm erg\ s^{-1}$ respectively. The black hole mass and the bolometric luminosity, both inferred from the \Pag\ broad component, amount to $M_{\rm BH} \sim 10^8\ \rm M_{\odot}$ and $L_{\rm bol} \sim 3 \times 10^{45}\ \rm erg\ s^{-1}$, respectively. Intriguingly, BiRD presents strict analogies with other two LRDs spectroscopically confirmed at cosmic noon, GN-28074 ("Rosetta Stone") at $z \sim 2.26$ and RUBIES-BLAGN-1 at $z \sim 3.1$. The blueshifted \Hei\ absorption detected in all three sources suggests that gas outflows may be common in LRDs. We derive a first estimate of the space density of LRDs at $z < 3$ based on JWST data, as a function of the bolometric luminosity and black hole mass. The space density $\Phi(L) = 4.0^{+4.0}_{-2.4} \times 10^{-6}$ Mpc$^{-3}$ dex$^{-1}$ is only a factor of $\sim 2-3$ lower than that of UV-selected quasars with comparable bolometric luminosity and redshift, meaning that the contribution of LRDs to the broader AGN population is also relevant at cosmic noon. A similar trend is also observed in terms of black hole masses. If, as suggested by recent theories, LRDs probe the very first and rapid growth of black hole seeds, our finding may suggest that the formation of black hole seeds remains efficient at least up to cosmic noon.}    
\keywords{quasars: supermassive black holes --- quasars: absorption lines --- galaxies: high-redshift --- galaxies: active --- galaxies: ISM}
\maketitle    

\section{Introduction} 
\label{sec:intro}
The James Webb Space Telescope (JWST; \citealt{gardner23}) has unveiled a population of active galactic nuclei (AGN) that was missed by previous selections \citep{harikane23, onoue23, ubler23, kocevski25, maiolino24, matthee24, juodzbalis25}. These AGN show moderately broad Balmer lines ($FWHM \gtrsim 1000-2000\ \rm km\ s^{-1}$), have bolometric luminosity $L_{\rm bol} \sim 10^{44}-10^{47}\ \rm erg\ s^{-1}$ and harbor massive black holes ($10^7-10^8\ \rm M_{\odot}$). Despite their AGN features, most of them stand out for lacking of both X-ray \citep{ananna24, mazzolari24b, yue24, maiolino25} and radio \citep{mazzolari24} emission , even if X-ray weakness was also found in high-$z$ red quasars not discovered by JWST \citep{fujimoto22, ma24}. When looked in their ultraviolet (UV) to near-infrared (NIR) images, a fraction of these JWST-discovered AGN ($\sim 10-30\%$; \citealt{hainline25}) show a point-like appearance, an increasingly red continuum, but also bright rest-frame UV colors \citep{greene24, labbe25}. Because of that, these AGN are commonly referred to as ‘little red dots’ (LRDs). LRDs are typically selected via NIRCam colors, looking for compact sources \citep{akins23, barro24, kokorev24, labbe25, perezgonz24}, or based on their rest-frame UV and optical slopes \citep{kocevski25}. However, the family of LRDs is much vaster, in the sense that not all LRDs show AGN features, namely broad lines, in their spectra. \citet{greene24} show that 60\% of LRDs in their spectroscopic sample have broad \Ha\ emission, while the remaining sources are either brown dwarfs or have an ambiguous identification. In addition, it should be kept in mind that the majority of JWST-discovered AGN does not have LRD colors and slopes and thus needs different selection techniques to be studied \citep{hainline25}. A characteristic signature of LRDs is the ‘v-shaped’ spectral energy distribution (SED), with the flex around the rest-frame wavelength 3645 \AA , i.e., the Balmer break wavelength. The prominent Balmer break was either interpreted as due to stellar population and/or scattered light \citep{labbe24} or to absorption by extremely dense gas (hydrogen density $n_{\rm H} \gtrsim 10^9\ \rm cm^{-3}$; \citealt{i&m25}). Absorption features are also often found in the broad Balmer lines and \Hei\ emission \citep{maiolino24, matthee24, juodzbalis24, kocevski25, lin24, ji25, wang25, deugenio25}. They are often blueshifted or redshifted, suggesting dense outflowing/inflowing material from the broad line region (BLR). Compared to low-redshift AGN, in which such kind of absorption is rare ($\sim 0.1\%$ detection rate; \citealt{i&m25}), Balmer absorption is much more common in JWST-discovered AGN and affects 10-20\% of them \citep{lin24}. This feature requires extremely high gas densities to significantly populate the $n = 2$ level in hydrogen atoms. For instance, \citet{juodzbalis24} infer $n_{\rm H} > 10^8\ \rm cm^{-3}$ in a broad line AGN. Recent studies proposed that the observed absorptions and the Balmer break result from the turbulent, dust-free, and highly dense atmosphere around the black hole (‘black hole star’ scenario; \citealt{naidu25, degraaff25}). It has also been suggested that the observed Balmer lines may be broadened either by resonant Balmer or electron scattering, implying an overestimate of the black hole masses by a factor $\sim 100$ if imputed to orbital motion only around the black hole \citep{naidu25, rusakov25}. However, these scenarios have been disputed by later studies \citep{juodzbalis24}. Detailed studies of absorption features in LRDs are thus pivotal for deciphering the elusive nature of these sources. 

Some studies attempted to quantify an LRD census as a function of redshift \citep{kokorev24, kocevski25, bisigello25, ma25}. \citet{kokorev24} report that their number density can be up to two orders of magnitude higher than UV-selected quasars with comparable magnitudes ($-22 < M_{UV} < -17$) at $6.5 < z < 8.5$. \citet{kocevski25} find that, while increasing at $z < 8$, the redshift distribution of LRDs in their sample shows a marked decline at $z \sim 4.5$. Moreover, LRDs are $\sim 1\ \rm dex$ more abundant than X-ray and UV-selected AGN at $z \sim 5-7$ with similar $M_{\rm UV}$. 
\citet{bisigello25} attempt to extend the LRD census to $z < 4$ using Euclid Quick Data Release. Unlike other studies, they found an increase in the number density of LRD candidates from $2.5 < z < 4$ to $z \sim 1.5-2.5$ and a decrease for lower redshifts. Compared to UV-selected quasars, the LRDs are less abundant up to UV magnitude $M_{UV} < -21$, while the two populations seem more compatible at fainter absolute magnitudes (where the quasar abundance is only extrapolated, though). However, a comparison of the two populations is not trivial because of the uncertainty in the absolute UV magnitude of LRDs. Finally, \citet{ma25} uses ground-selected LRDs at $2 \lesssim z \lesssim 4$ finding a drop in their number density at $z \sim 4$ that is qualitatively consistent with that found by \citet{kocevski25}. However, their results are $1\ \rm dex$ higher than the predictions at $z \sim 2$ by \citet{inayoshi25}, based on JWST-discovered LRDs. In this uncertain picture, studying LRDs at $z < 4$ is thus necessary to firmly assess their abundance and their relative contribution to the broader AGN population.

To date, there are only a few spectroscopically confirmed LRDs or possible analogous \citep{stepney24, rinaldi25} at the so-called ‘cosmic noon’ \citep{juodzbalis24, hviding25, ma25, wang25, zhuang25}, i.e., $z\sim 2-3$, when the cosmic star formation and black hole accretion rate density peak \citep{madau&dick, delvecchio14, vito18}. 
Two of them, in particular, have been studied in detail: \citet{juodzbalis24} discovered a broad-line AGN at $z = 2.26$ in the GOODS-N field using data from the JADES (JWST Advanced Deep Extragalactic Survey) program. The object, named GN-28074 (and dubbed the “Rosetta Stone”), in addition to prominent broad Balmer and \Hei\ lines, shows clear absorption features in \Ha , \Hb , and \Hei , while its colors pass the selection of LRDs defined by \citet{kocevski25}, when redshifted at $z = 7$. 
Similarly, \citet{wang25} found a LRD at $z = 3.1$ as part of the RUBIES (Red Unknowns: Bright Infrared Extragalactic Survey; \citealt{degraaff24}) program. The source, dubbed RUBIES-BLAGN-1, shows multiple broad lines, \Hei\ absorption, a blue UV-continuum, and a Balmer break, similar to other broad-line AGN that are also LRDs.

Here we report the discovery of a bright LRD, named BiRD ('big red dot'), at $z \sim 2.33$ in the \textit{J1030 field}, a well studied region around the $z \sim 6.3$ quasar SDSS J1030+0524\footnote{\url{http://j1030-field.oas.inaf.it/}}. The field was imaged from the ground in the optical and NIR (LBT/LBC, CFHT/WIRCam) and also observed by \textit{Spitzer}/IRAC \citep{annunziatella18} and MIPS, and is part of the MUSYC survey \citep{gawiser06}. At millimeter wavelength, it was covered by \textit{Az}TEC \citep{zeballos18}. The central $1' \times 1'$ region was observed by HST/ACS, HST/WCF3, VLT/MUSE and ALMA \citep{stiavelli05, damato20, mignoli20}. Furthermore, it is among the deepest X-ray extragalactic fields \citep{nanni20}, and it has been covered by deep 1.4 GHz JVLA data \citep{damato22}. The field around the quasar was also targeted by the JWST GTO (Guaranteed Time Observations) program EIGER (Emission-line galaxies and Intergalactic Gas in the Epoch of Reionization), which covered it with the Near Infrared Camera (NIRCam) imaging and wide field slitless spectroscopy (WFSS). 

In this work we compare our target with the LRDs at cosmic noon studied by \citet{juodzbalis24} and \citet{wang25} for which a detailed analysis of the \Pag\ and \Hei\ lines, as well as a broad-band SED and black hole masses, have been published, allowing for a complete comparison with the measurements that we perform on our source. 

The paper is organized as follows: in Sect.~\ref{sec:datared} we describe the data reduction and in Sect.~\ref{sec:sel} the target selection. Sect.~\ref{sec:analysis} deals with the spectral and SED analysis, while in Sect.~\ref{sec:res} we report the results. Finally, we discuss our findings in Sect~\ref{sec:disc} and we summarize our conclusions in Sect~\ref{sec:concl}.

In this work we adopt a $\Lambda$CDM cosmology with $\Omega_{\Lambda} = 0.7$, $\Omega_{\rm M} = 0.3$ and a Hubble constant $H_0 = 70\ \rm km\ s^{-1}\ Mpc^{-1}$. The reported errors correspond to the statistical uncertainty of 68\% (1$\sigma$), if not specified differently. 
\section{Data reduction}
\label{sec:datared}
\begin{figure}
\begin{center}
\includegraphics[width=0.52\textwidth]{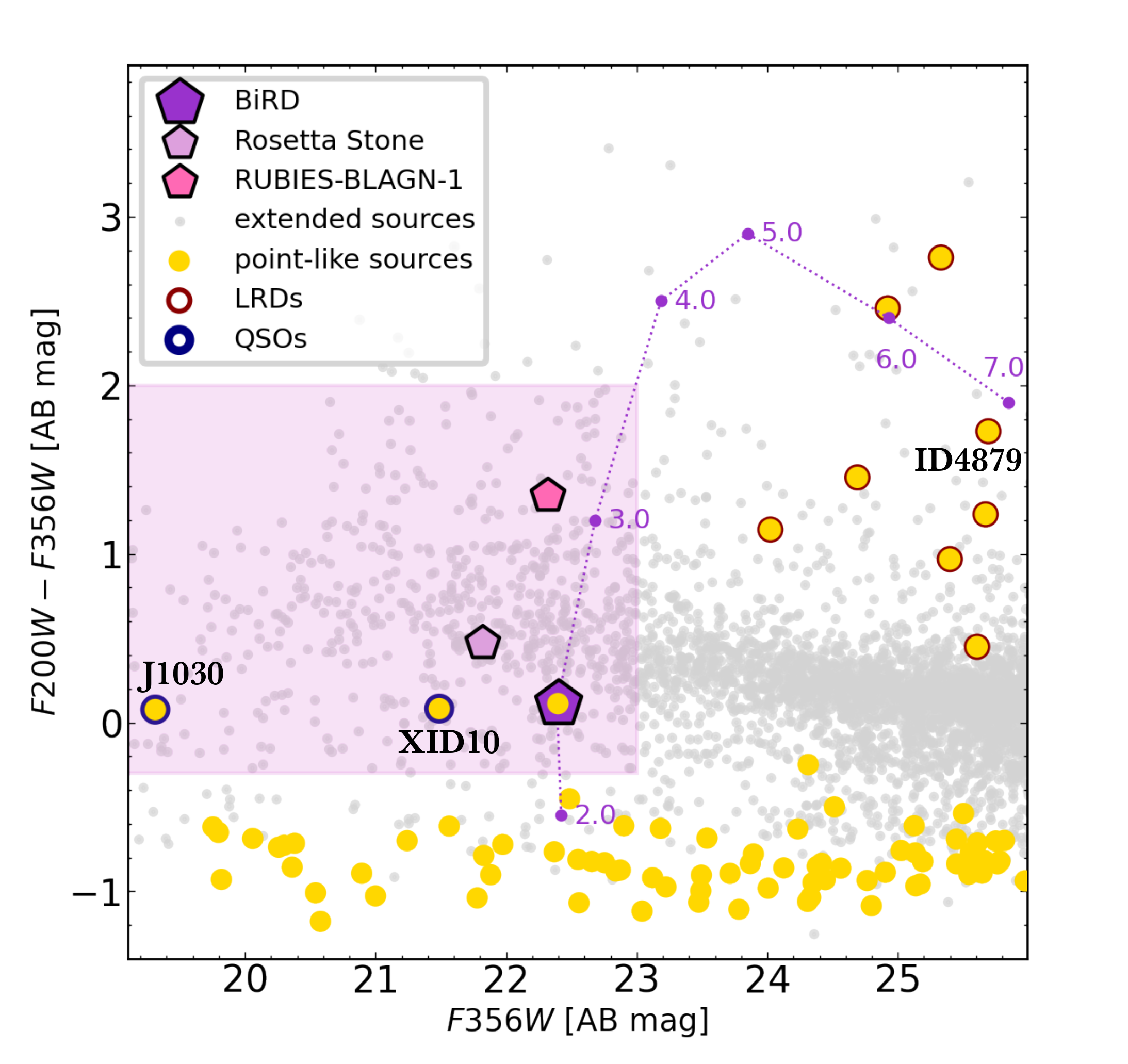}\\
\end{center}
\caption{$F200W-F356W$ color vs. $F356W$ magnitude distribution of JWST-detected sources in the J1030 field. The gray dots represent the extended sources, while point-like sources are shown as yellow filled circles. Stars align along a horizontal sequence at $F200W-F356W\sim -0.9$. Faint ($F356W>24$) and red ($F200W-F356W >0$) point-like objects (filled circles with red contour) are mostly LRD candidates at $z>4$ (e.g., ID4879 at $z=5.95$; see text for details). The three bright ($F356W<23$), point-like objects above the stellar sequence are two standard, X-ray detected broad line AGN (J1030 and XID10; filled circles with blue contours) and the BiRD (purple pentagon; see text for details). The dotted line shows the track followed by the BiRD if moved at different redshifts as labeled from $z = 2$ to $z = 7$.  The smaller pentagons show for comparison the position of two spectroscopically confirmed LRDs at cosmic noon, discovered with JWST, namely the Rosetta Stone at $z=2.26$ (plum; \citealt{juodzbalis24}) and RUBIES-BLAGN-1 at $z=3.10$ (pink; \citealt{wang25}). The colored box shows the selection region used to compute the space density of LRDs at cosmic noon (see Section~\ref{sub:spaceden}).}
\label{fig_dplot}
\end{figure}

We used the archival data of the 1243 GTO program EIGER (PI: S. J. Lilly). Briefly, this program targets the fields of six quasars (J0100+2802, J0148+0600, J1030+0524, J159-02, J1120+0641, and J1148+5251) at redshift $5.9 < z < 7$ with NIRCam imaging (in the $F115W$, $F200W$, and $F356W$ filters) and NIRCam/WFSS ($F356W$ filter and $R$ grism); see \citet{kashino23} for further details.

In our analysis, we focused on the J1030+0524 field (hereafter, J1030). The NIRCam images form a mosaic of $\sim 6.5' \times 3.4'$ around the quasar. The exposure time varies across the field and reaches in the central part the value of $\sim$17~ks, $\sim$20~ks, and\ $\sim$6\ ks for the $F115W$, $F200W$, and $F356W$ filters, respectively. We reduced the imaging data with the 1.14.0 version (\texttt{jwst\_1293.pmap}) of the JWST pipeline \citep{bushouse}. The reduction consists of three stages. In short, we processed the raw data (\textit{uncal} files) through the \textit{Stage1} and \textit{Stage2} of the pipeline. We subtracted the correlated readout noise ($1/f$ noise), the background, and scattered light (\textit{"wisps"}) from the calibrated exposures (\textit{cal} files). We registered the $F356W$ mosaic on the multiband $Ks$-selected catalog of J1030 (Mignoli et al., in prep.) and then registered the $F115W$ and $F200W$ mosaics on the $F356W$ one.
More details about the data reduction will be reported in future work.

Depending on the wavelength, the WFSS data of J1030 cover an area $\sim 17 \ \rm arcmin^2$ at $3.1\ \mu m$ and $\sim 25 \ \rm arcmin^2$ at $3.95\ \mu m$ (see \citealt{kashino23}). The observed spectral range is $3.1-4.0\ \mu m$ and the resolving power $R = \lambda/ \Delta \lambda$ is $\sim 1500$ at the central wavelength. We processed the raw exposures through the \textit{Stage1} of the pipeline. The products of this stage (\textit{rate} files) were reduced with the routine\footnote{\url{https://github.com/fengwusun/nircam_grism/}} developed by Fengwu Sun (see \citealt{sun23}). We applied the flat-field, subtracted the background and the $1/f$ noise, and registered the astrometry of the grism exposures on the short-wavelength (SW) filters (i.e., $F115W$ and $F200W$). We used the BiRD position (R.A.: 157.57507, Dec.: 5.37915) in the JWST catalog of J1030 (see Sect~\ref{sub:cat}) as prior for the 2D spectral extraction. More details will be given in a future work.

\begin{table}
\caption{Photometry of the BiRD}  
\label{tab:phot}     
\centering                                
\scalebox{0.8}{
\begin{tabular}{c c} 
\hline\hline
Band & mag \\ & [AB]  \\
\hline  
B-CTIO & 26.40 $\pm$ 0.20 \\
g-LBC & 26.73 $\pm$ 0.21 \\
r-LBC & 26.09 $\pm$ 0.11 \\
i-LBC & 25.75 $\pm$ 0.14 \\
z-LBC & 25.82 $\pm$ 0.21 \\
Y-WIRCam &$>$25.3 \\
J-WIRCam &$>$24.9 \\
K-WIRCam & 22.15 $\pm$ 0.08 \\
F115W & 25.41 $\pm$ 0.07 \\
F200W & 22.49 $\pm$ 0.04 \\
F356W & 22.37 $\pm$ 0.06 \\
3.6-IRAC &22.56 $\pm$ 0.13\\
4.5-IRAC &23.02 $\pm$ 0.22\\
5.8-IRAC &$>$22.0 \\
8.0-IRAC &$>$21.7 \\
21$\mu$m-MIPS &$>$19.6 \\
[1mm]
\hline 
\end{tabular}}
\tablefoot{The quoted magnitudes are the aperture-corrected ones that agree, within the photometric errors, with the total (AUTO) magnitudes computed by SExtractor, suggesting a compact nature of the target.}
\end{table}

\section{Target selection}
\label{sec:sel}
\subsection{JWST catalog}
\label{sub:cat}
\begin{figure*}
\begin{center}
\includegraphics[width=1.0\textwidth]{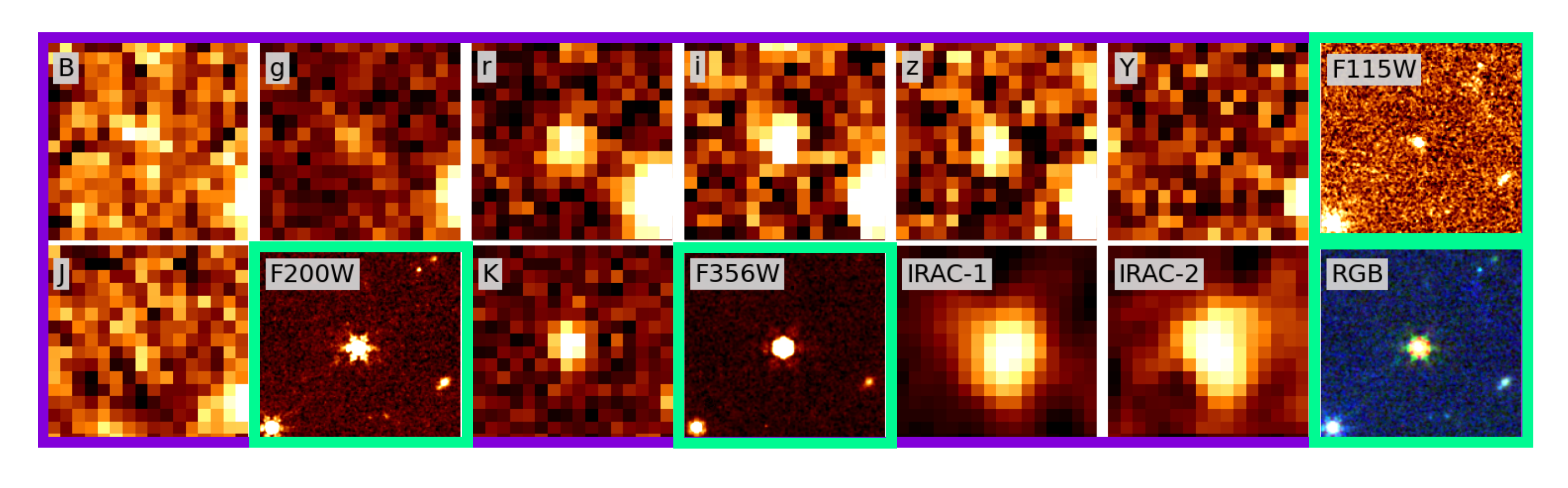}\\
\end{center}
\caption{Multiwavelength cutouts of BiRD. The images ($5\arcsec \times 5\arcsec$ each) were obtained with CTIO (B; MUSYC survey; \citealt{gawiser06}), LBT/LBC (g,r,i,z), CFHT/WIRCAM (Y,J,K), \textit{JWST}/NIRCam (F115W, F200W, F356W), and \textit{Spitzer}/IRAC (CH1 and CH2; \citealt{annunziatella18}). The JWST images are enclosed in green squares. The RGB image includes the NIRCam filters only.}
\label{fig_cutouts}
\end{figure*}
Based on the NIRCam images, we build the JWST catalog of J1030. We will describe the catalog in detail in a future work while here we briefly summarize the main information. For photometry on JWST NIRCam images, we use SExtractor \citep{bertin&arnouts96} with parameters DETECT\_MINAREA = 12 and DETECT\_THRESH = 2.5, adopting relatively conservative detection settings, as the primary purpose of the catalog is the extraction of slitless spectra. The analysis is performed in dual-image mode, using the \textit{F356W} band as the detection image, since the WFSS spectroscopic data were acquired in this band. We measure fluxes in the \textit{F115W} and \textit{F200W} filters using circular apertures of various radii, adopting a reference diameter of 0.18\arcsec.
Given the relatively small differences between the PSFs in the three bands, the ancillary images were not convolved to match the PSF of the \textit{F356W} detection image. Instead, point-source aperture corrections were applied to compute accurate photometric colors. We measured and applied aperture corrections of 0.60, 0.28, and 0.28 for the \textit{F356W}, \textit{F200W}, and \textit{F115W} filters, respectively. This approach optimizes the use of high signal-to-noise pixels and enhances the ability to detect faint counterparts in the two bluer filters without degrading their PSF quality. 

The final photometric catalog includes 10,692 entries. Since the J1030 field is not uniformly covered by the NIRCam observations, we computed the 3$\sigma$ magnitude limit per pixel for each observed band using the noise maps produced during data reduction. For forced photometry in the \textit{F200W} and \textit{F115W} bands, measured at the positions of sources detected in the \textit{F356W} image, we compared the aperture magnitude with the corresponding detection limit. If the measured magnitude was fainter than the 3$\sigma$ threshold, the limiting value was adopted instead. As a result, 543 sources in \textit{F200W} (5\%) and 1,318 sources in \textit{F115W} (12\%) are flagged as upper limits in flux relative to the \textit{F356W}-detected sources.
Due to dithering, the mosaic does not have uniform exposure coverage, resulting in spatial variations in the photometric limiting magnitude. However, in a significant fraction of the image ($>$80\%), the photometric catalog reaches a 5$\sigma$-limiting magnitude of \textit{F356W}~=~28.5.
Finally, to achieve a robust morphological selection of the sources in the photometric catalog, we employed two techniques. The first relies on the simple ratio between the total flux of an object and its flux within a fixed aperture. This method efficiently selects point-like sources down to a magnitude limit of \textit{F356W}~$<$~26, where the locus of faint and compact galaxies intersects that of stellar objects. We further analyzed the list of candidate point-like sources selected above using the IRAF task \textit{psfmeasure}, which estimates the PSF width of the targets by evaluating the enclosed flux as a function of the radial profile. Only two of the previously selected point-like sources did not pass the subsequent PSF analysis. The two-step morphological analysis yielded a robust list of 88 objects classified as point-like, out of a total of 3404 sources in the photometric catalog down to a magnitude of \textit{F356W}~=~26.

\subsection{Discovery plot}
\label{sub:anc}
\begin{figure*}
\begin{center}
\includegraphics[width=1\textwidth]{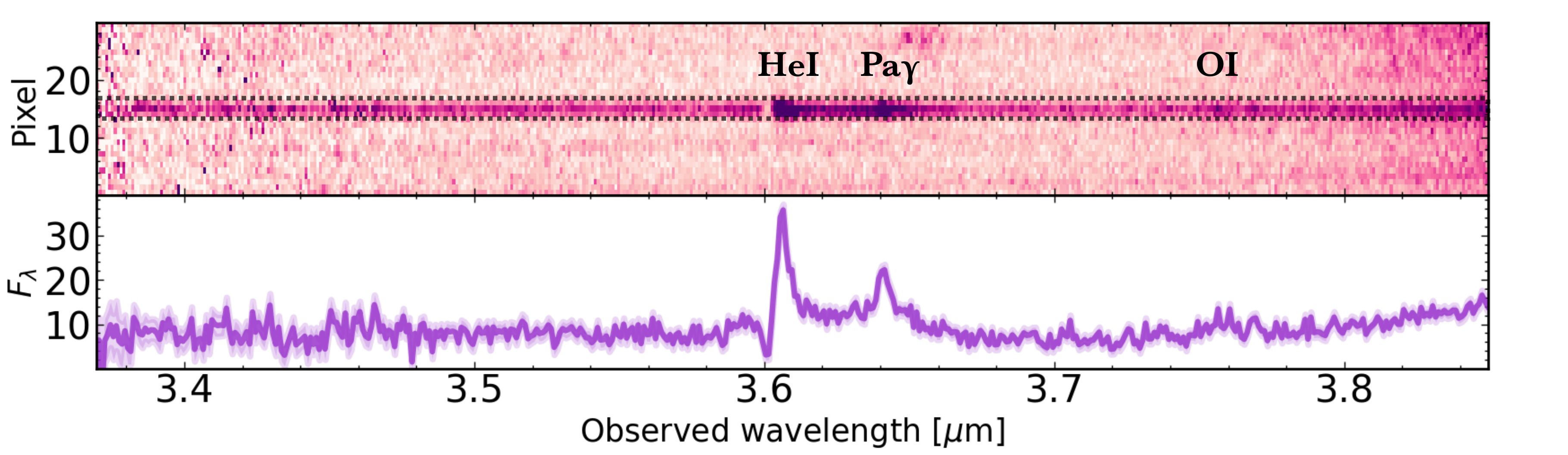}\\
\end{center}
\caption{Two-dimensional and one-dimensional spectrum of BiRD (NIRCam/WFSS). The dotted box indicates the 5-pixel aperture used to extract the one-dimensional spectrum. We see clear emission from \Hei , \Pag\ and \Oip\ lines, superimposed on the continuum. The contamination from a bright source is visible at the red end of the spectrum. The flux density $F_{\lambda}$ is expressed in units of $10^{-16}\ \rm erg\ s^{-1}\ cm^{-2} \mu m^{-1}$. The $1\sigma$ error spectrum is reported as a purple shaded area.}
\label{fig_2dspec}
\end{figure*}

We selected our target as a bright outlier in the color vs. magnitude, $F200W-F356W$ vs. $F356W$ mag, distribution of point-like sources (yellow points in Fig.~\ref{fig_dplot}). As shown in Fig.~\ref{fig_dplot}, point-like sources mainly separate into two distinct groups: stars align on a horizontal sequence centered at $F200W-F356W\sim -0.9$, whereas the group of red, $F200W-F356W >0$, and faint, $F356W>24$, point-like sources is instead mainly populated by high-$z$ LRD candidates. Besides their red color and compactness, we have limited information to classify this group as LRDs. For instance, because of their faintness, these objects are mostly undetected in the optical imaging of the J1030 field, even in the region with HST/ACS coverage. However, we remark that in the few cases where a spectroscopic confirmation is available from the WFSS spectrum, an inflection point consistent with the wavelength of the Balmer break is observed even with the limited photometric coverage. We highlight in Fig.~\ref{fig_dplot} one example at $z=5.95$ (ID4879), where the detection of \Hb\ and of the \Oiii\ $\lambda \lambda\ 4959,5007$ doublet allows for an unambiguous redshift determination. A full discussion of the population of LRD candidates in the J1030 field is beyond the scope of this paper and will be presented elsewhere.\\
\indent We inspected the three objects with $F356W<23$ and $F200W-F356W \sim 0$,  as they appear redder than stars and brighter than LRDs. The brightest two are luminous, X-ray detected, broad line quasars: namely J1030 at $z=6.31$ and XID10 at $z=2.74$ \citep{marchesi21}, with $F356W=19.3$ and 21.5, respectively. The third object, with $F356W\sim 22.4$, is not detected in the X-rays, nor in the radio band (see Sect.~\ref{sub:xrp}), and was not previously identified spectroscopically. 
In Table~\ref{tab:phot} we present all the available photometry for this object, which was also previously detected in our ground-based photometry (Mignoli et al., in prep.), while in Fig.~\ref{fig_cutouts} we show the cutouts.
The corresponding WFSS two-dimensional spectrum shows two emission lines superimposed on continuum emission (Fig.~\ref{fig_2dspec}). Unfortunately, in the red part ($\lambda >$ 3.8 $\mu$m) of the frame, a bright source contaminates the target's spectrum. We optimally extracted the spectrum over 5 pixels and identified the two lines as the \Hei\ $\lambda\ 10830$ emission and Paschen~$\gamma$ (\Pag , hereafter), obtaining an initial redshift estimate of $z \sim 2.33$. Once the redshift was estimated based on the two brightest lines in the spectrum, a faint emission line at 37595 \AA\ was also identified and attributed to the \Oi\ $\lambda\ 11290$ line. We named the source BiRD ('big red dot') because it stands out in the NIRCam images when compared to the other, much fainter, LRD candidates at higher redshifts in the field.

\section{Analysis}
\label{sec:analysis}
\subsection{Spectral line fitting}
\label{sub:fit}
\begin{figure}
\begin{center}
\includegraphics[width=0.45\textwidth]{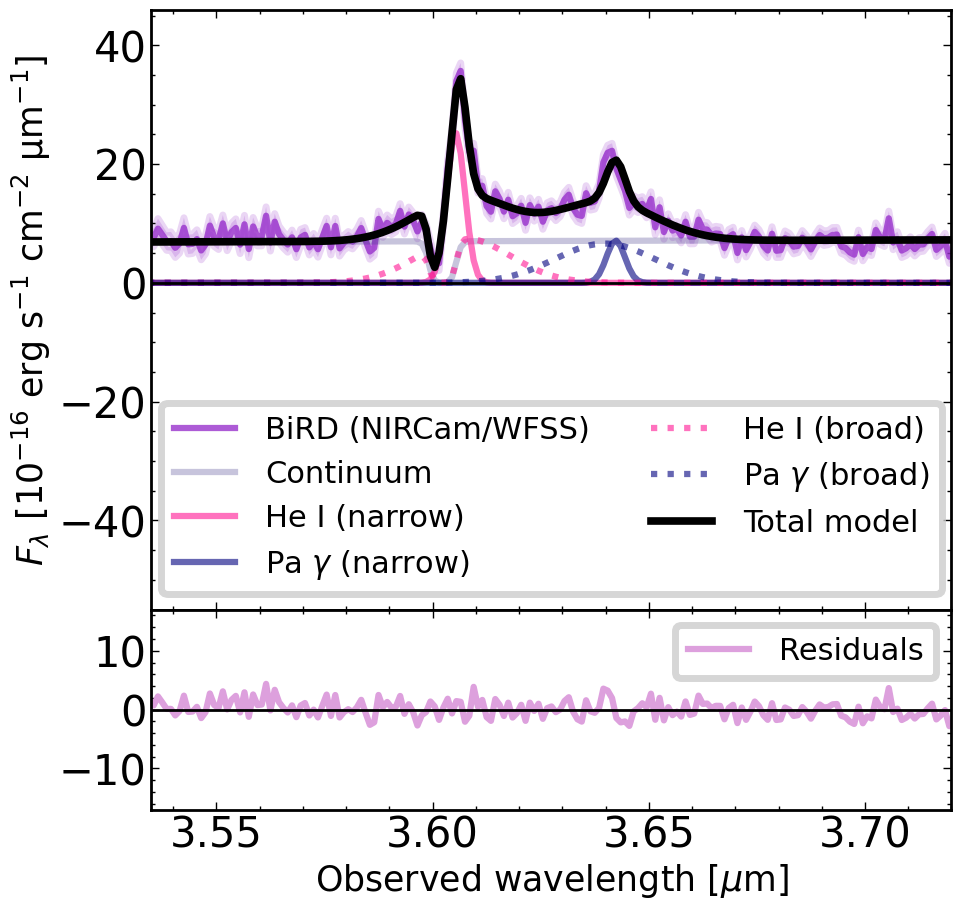}\\
\end{center}
\caption{Continuum and line fit of BiRD. The NIRCam/WFSS spectrum is shown in purple with the corresponding errors (shaded area). We modeled both the \Hei\ (pink) and the \Pag\ (violet) emission with a narrow and a broad Gaussian component. An absorption term was applied to both continuum emission and broad \Hei\ line to account for the feature at $3.60\ \mu \rm m$}. We show the fit residuals in the bottom panel (lilac).
\label{fig_fit}
\end{figure}

We fit the BiRD spectrum using a Markov chain Monte Carlo (MCMC) method with the Python package \texttt{emcee} \citep{foreman13}. We used a linear continuum and four Gaussian components to reproduce the line emission. We modeled both \Hei\ and \Pag\ using a narrow and a broad component. We tied the narrow components to have the same redshift, which we assumed to be the systemic one, and the same width, as we expect that they are both emitted from the narrow line region (NLR). We applied an absorption term to the broad \Hei\ line and to the continuum emission to account for a clear absorption feature at $3.6\ \rm \mu m$ (see Fig.~\ref{fig_fit}). Starting from Eq. (9.2) of \citet{draine11}, we modeled the observed, attenuated flux $I(\lambda)$ of these two components in the wavelength space as:
\begin{equation}
    I(\lambda) = I_0(\lambda)\ e^{-\tau(\lambda)}
\end{equation}
where $I_0(\lambda)$ is the intrinsic, unabsorbed continuum or the \Hei\ unattenuated broad line flux and $\tau(\lambda)$ is the optical depth near line center. The latter is parametrized as (see \citealt{draine11}, Eq. 9.7): 
\begin{equation}
    \tau(\lambda) = \tau_0\ e^{-0.5 \left(\frac{c(1-\lambda_0/\lambda)}{\sigma_v}\right)^2}
\end{equation}
where $\tau_0$ is the optical depth at the line center, $\lambda_0$ is the absorber central wavelength, $\sigma_v$ is the velocity dispersion and $c$ the speed of light (both in \kms). We are assuming here that the absorber affects the radiation from the accretion disk and the broad line region (BLR), i.e., it is located in the BLR or just outside it, to compare with other works (see, for instance, \citealt{deugenio25, i&m25, ji25, juodzbalis24}). However, we note that, based on our data only, the absorber location is unconstrained. We thus discuss a different location of the absorber in Sect.~\ref{sub:abs}, and specifically how applying the absorption term also to the narrow \Hei\ component affects the column density of the absorber.  

We used uniform priors for all parameters. For the narrow and broad components, we allowed the line width to be less than 1000 and 10000 \kms\ respectively. We estimated the best values as the median of 10000 realizations and the errors as the 16th and 84th percentiles. We derived the quantities presented in Table~\ref{tab:fit} from the fitted parameters, while the fit is shown in Fig.~\ref{fig_fit}. For the absorber, we report the derived quantities in Table~\ref{tab:abs} (see Sect.~\ref{sub:abs} for the column density). We take into account the instrumental broadening by subtracting in quadrature the instrumental $FWHM$, which is $FWHM_{\rm instr} \sim 83\ \rm km\ s^{-1}$ at our wavelengths, from the $FWHM$ of the fitted lines. The correction was applied only to the narrow components and for the absorber, as it is irrelevant for the broad components.

We estimate from the narrow components a systemic redshift $z = 2.3290_{-0.0002}^{+0.0003}$. The velocity shifts of the broad \Hei\ and \Pag\ are essentially unconstrained by our fit, while the FWHM of both lines exceed 2000 \kms .
The absorber is blueshifted by $-830_{-148}^{+131}$ \kms compared to the \Hei\ narrow component (see Table~\ref{tab:abs}). The fitted quantities are independent of the adopted assumptions. We checked indeed that by performing a free fit, i.e., without any tied parameters, or tying the line width and centroid also of the broad components, we obtain values that are consistent within the errors with the quantities reported in Table~\ref{tab:fit} and Table~\ref{tab:abs}.
\begin{table}
\caption{Quantities derived from the line fit}  
\label{tab:fit}      
\centering                                
\scalebox{0.8}{
\begin{tabular}{c c c c}  
\hline\hline 
Line & Flux & $\Delta v$ & FWHM \\[1mm] 
     & ($10^{-16}\ \rm erg\ s^{-1}\ cm^{-2}$) & (\kms) & (\kms) \\
\hline  
\Hei\ (narrow) & $13_{-2}^{+2}$ & $\_$ & $367_{-34}^{+41}(*)$ \\[1mm] 
\Pag\ (narrow) & $3.8_{-0.8}^{+0.8}$ & $\_$ & $367_{-34}^{+41}(*)$ \\[1mm]
\Hei\ (broad) & $21_{-3}^{+3}$ & $564_{-433}^{+677}$ & $2073_{-337}^{+572}$ \\[1mm] 
\Pag\ (broad) & $19_{-3}^{+3}$ & $-691_{-488}^{+498}$ & $2237_{-309}^{+342}$ \\[1mm]
\hline 
\end{tabular}}
\tablefoot{The errors are computed from the 16th and 84th percentiles of the fitted parameters in the MCMC fit. The $(*)$ means that the quantity was corrected for the instrumental broadening.}
\end{table}

\begin{table}
\caption{Properties of the absorber}  
\label{tab:abs}      
\centering                                
\scalebox{0.8}{
\begin{tabular}{c c c c c c}  
\hline\hline  
Line & $\tau_0$ & $\Delta v$ & FWHM & $N$ \\[1mm] 
     & & (\kms) & (\kms) & ($\rm cm^{-2}$) \\
\hline  
\Hei\ (absorber) & $5_{-3}^{+9}$ & $-830_{-148}^{+131}$ & $242_{-68}^{+77} (*)$ & $1.2_{-0.4}^{+1.5} \times 10^{14}$ \\[1mm]
\hline 
\end{tabular}}
\tablefoot{The errors are computed from the 16th and 84th percentiles of the fitted parameters in the MCMC fit. The $(*)$ means that the quantity was corrected for the instrumental broadening.}
\end{table}

\subsection{SED analysis}
\label{sub:sed}

With the source redshift in hand, we inspected its rest-frame broadband SED by combining the JWST photometry with optical, UV, and NIR photometry from the ground (the BiRD falls outside the region covered by HST/ACS and WFC3 observations), and with Spitzer NIR and mid-IR photometry. For the derivation of ground-based magnitudes and IRAC-CH1 and -CH2 magnitudes (3.6 and 4.5 $\mu$m, respectively), we refer the reader to Mignoli et al., in prep. (see also \citealt{morselli14,annunziatella18}). As for IRAC-CH3 and -CH4 magnitudes (5.8 and 8.0 $\mu$m, respectively) and MIPS 21 $\mu$m magnitude, we downloaded the available mosaics from the Spitzer Enhanced Imaging Products (SEIP) archive\footnote{https://irsa.ipac.caltech.edu/data/SPITZER/Enhanced/SEIP/} and performed photometric measurements based on an updated version of the PhoEBO pipeline \citep{gentile24}.
This pipeline uses high-resolution $Ks$-band images to define prior positions and shapes of sources, which are then used to separate blended objects in the lower-resolution IRAC-CH3, IRAC-CH4, and MIPS 24 $\mu$m bands. The priors were extracted with the Python implementation of Source Extractor \citep{barbary16}. To model how these sources would appear in the lower-resolution images, we convolved the priors with matching kernels built from the PSFs of each band, and the model best-fit fluxes were derived by minimizing the residuals over the observed data (see \citealt{gentile24}).
The BiRD is not detected in the IRAC-CH3, IRAC-CH4 and MIPS 24 $\mu$m maps. The corresponding 3$\sigma$ limiting magnitudes are reported in Table~\ref{tab:phot}.

As shown in Fig.~\ref{fig_sed}, the rest-frame broadband SED of the BiRD shows an inflection point close to the 3645 \AA\ Balmer break, reminiscent of the 'v'-shaped SED of LRDs. We fitted power-law spectra of the form $f_\lambda\propto \lambda^\beta$ to the photometric data-points blueward and redward of the Balmer break, and determined the UV and optical rest spectral slopes, respectively. We find $\beta_{\rm UV,\ rest}=-0.80\pm0.16$ for the UV slope. For the optical slope, we find $\beta_{\rm opt,\ rest}=2.74\pm0.23$ when limiting the fit to the data-points below 7000 \AA\ rest-frame, i.e., where the SED peaks. As H$\alpha$ emission can contaminate photometric measurements around this peak, at 2 $\mu$m in the observed frame, we also performed a fit conservatively ignoring the JWST $F200W$ and the WIRCAM K-band photometry, and just used the JWST F115W and F356W photometry, finding $\beta_{\rm opt,\ rest}=0.22\pm0.16$. We note that this second approach is very conservative since, based on what is observed in the 'Rosetta Stone' \citep{juodzbalis24}, removing H$\alpha$ emission should not reduce the observed K-band flux by more than $\sim 30\%$. In either way, our measured slopes satisfy the selection criteria of $\beta_{\rm UV,\ rest}<-0.37$  and $\beta_{\rm opt,\ rest}>0$ adopted by \citet{kocevski25} to select LRDs at all redshifts. Therefore, the BiRD stands out as a spectroscopically confirmed LRD at cosmic noon.

In Fig.~\ref{fig_sed} we also compare the rest-frame SED of the BiRD with that of the Rosetta Stone at $z=2.26$ \citep{juodzbalis24} and RUBIES-BLAGN-1 \citep{wang25} at $z=3.1$, and with the median rest-frame SED of 675 LRDs at all redshifts derived by \citet{casey24}. Except for a downturn above $\sim1\ \mu$m rest-frame, the BiRD has a very similar SED to the Rosetta Stone and RUBIES-BLAGN-1. In addition, because of their relatively low redshift, they are a factor of 10-30 brighter than the average population of LRDs. 

\section{Results}
\label{sec:res}
\subsection{Bolometric luminosity, X-ray and radio properties}
\label{sub:xrp}
Most spectroscopically identified LRDs display broad Balmer lines in their spectra, which are commonly used to estimate their bolometric luminosity and black hole mass, assuming standard scaling relations for quasars (see Sect.~\ref{sub:bh}). In particular, the extinction-corrected luminosity of the broad \Ha\ component can be converted into bolometric AGN luminosity using the relation of \citet{stern12}:

\begin{equation}
L_{\rm bol} = 130^{\times2.4}_{\div 2.4} \; L_{\rm H\alpha,broad}    
\end{equation}

In the absence of direct spectroscopic coverage at $\sim 6500$ \AA\ rest-frame, we estimated the luminosity of the broad \Ha\ line starting from the broad \Pag\ component and assuming Case B in the theory of recombination-line radiation. This gives $L_{\rm H\alpha,broad} = 31.7 \times L_{\rm Pa\gamma,broad} \sim 2.6 \times 10^{43}\ \ergs$, and, in turn,  $L_{\rm bol}\sim 3\times10^{45}\ \ergs$ ($1.2-7.0\times10^{45}\ \ergs$ when accounting for the dispersion in the \citealt{stern12} relation).
We note that, even if Case B may not always apply in the high-density AGN BLR \citep{Ilic12}, it seems to work well for the Rosetta Stone: as a matter of fact, the luminosity of the broad Pa$\gamma$ line measured in the Rosetta Stone aligns very well with the expectations of case B recombination once the luminosities of its broad H$\alpha$ and H$\beta$ lines are corrected for extinction using the observed Balmer decrement. We discuss further implications of adopting case B recombination in Sect.~\ref{sub:sed-discuss}.

Based on the estimated bolometric luminosity, we verified whether the BiRD, similarly to most LRDs, is severely underluminous in the X-rays compared to standard AGN. 
The BiRD is not detected in the deep, $500\rm \ ks$ Chandra exposure of the J1030 field \citep{nanni20,marchesi21}. It is located at 3.25 arcmin from the exposure-weighted aimpoint of the Chandra mosaic, where the 90\% encircled energy radius is 1.5\arcsec\ at 1.5 keV and 2.5\arcsec\ at 6.4 keV.
To estimate approximate $3\sigma$ upper limits to source counts, we considered background counts extracted within annuli of inner radii 1.5\arcsec, 2.0\arcsec\ and 2.5\arcsec, for the 0.5-7 keV (full, FB), 0.5-2 keV (soft, SB), 2-7 keV (hard, HB) band, respectively, and outer radii of 10\arcsec\ for all X-ray bands, and then computed the binomial no-source probability \citep{weisskopf07,vito19,nanni20}.
We obtained $\sim$3$\sigma$ limits of 5, 3, and 6 counts in the FB, SB, and HB, respectively, that we then converted into limits to the source count rates after correcting for the assumed aperture and for the small loss of effective area at the BiRD's position. Finally, assuming a standard power-law spectrum with a photon index of $\Gamma=1.9$ \citep{peca21,signorini23} and no intrinsic absorption, we obtained $3\sigma$ upper limits of $1.7\times 10^{-16}, 7.1\times 10^{-17}, 3.3\times 10^{-16}\ \cgs$ to the source flux in the FB, SB, and HB, respectively. The soft band flux returns the most stringent constraint to the 2-10 keV rest-frame luminosity of $L_{\rm 2-10\ keV,rest}<3.7\times 10^{42}\ \ergs$. This, in turn, implies a bolometric correction of $L_{\rm bol}/L_{\rm 2-10\ keV,rest} > 800$, which is at least 30 times larger than the average expected value for standard AGN (see Fig.\ref{fig_xray}) and therefore calls for either very heavy suppression or intrinsically weak X-ray emission \citep{maiolino25}. 

The BiRD is also undetected in the deep radio image of the J1030 field, which reaches a {\it rms} of 2.5 $\mu$Jy at 1.4 GHz \citep{damato22}. The BiRD falls in the region of maximum radio sensitivity, and its non-detection corresponds to a $3\sigma$ limit to the rest frame 1.4~GHz luminosity of $L_{\rm 1.4\ GHz}<3\times 10^{39}\ \ergs$, assuming a standard spectral slope of
$\alpha=-0.7$ (with $S_{\nu}\propto \nu^{\alpha}$). We estimated the expected radio luminosity of BiRD assuming that it is a standard AGN and using literature calibrations \citep{panessa19, panessa15, damato22}. We considered the correlation between intrinsic 2-10 keV and 1.4 GHz luminosity found by \citet{damato22} for non-jetted (radio-quiet) AGN. We derived the intrinsic expected 2-10 keV X-ray luminosity from the bolometric one using the correlation from \citet{duras20}, obtaining $L_{\rm 1.4\ GHz,exp}=6.7\times 10^{39}\ \ergs$. This would increase to $L_{\rm 1.4\ GHz, exp}=10^{40}\ \ergs$ if instead the intrinsic X-ray luminosity was computed using the $\rm L_{H\alpha}-L_{2-10\ keV, rest}$ relation from \citet{jin12}.
The radio luminosity of the BiRD is therefore at least $0.3-0.5$ dex lower than expected on the basis of average correlations for standard AGN. However, because of the systematics and scatter in the adopted relations, deeper radio data would be needed to confirm this difference and assess its significance.
\begin{figure}
\begin{center}
\includegraphics[width=0.52\textwidth]{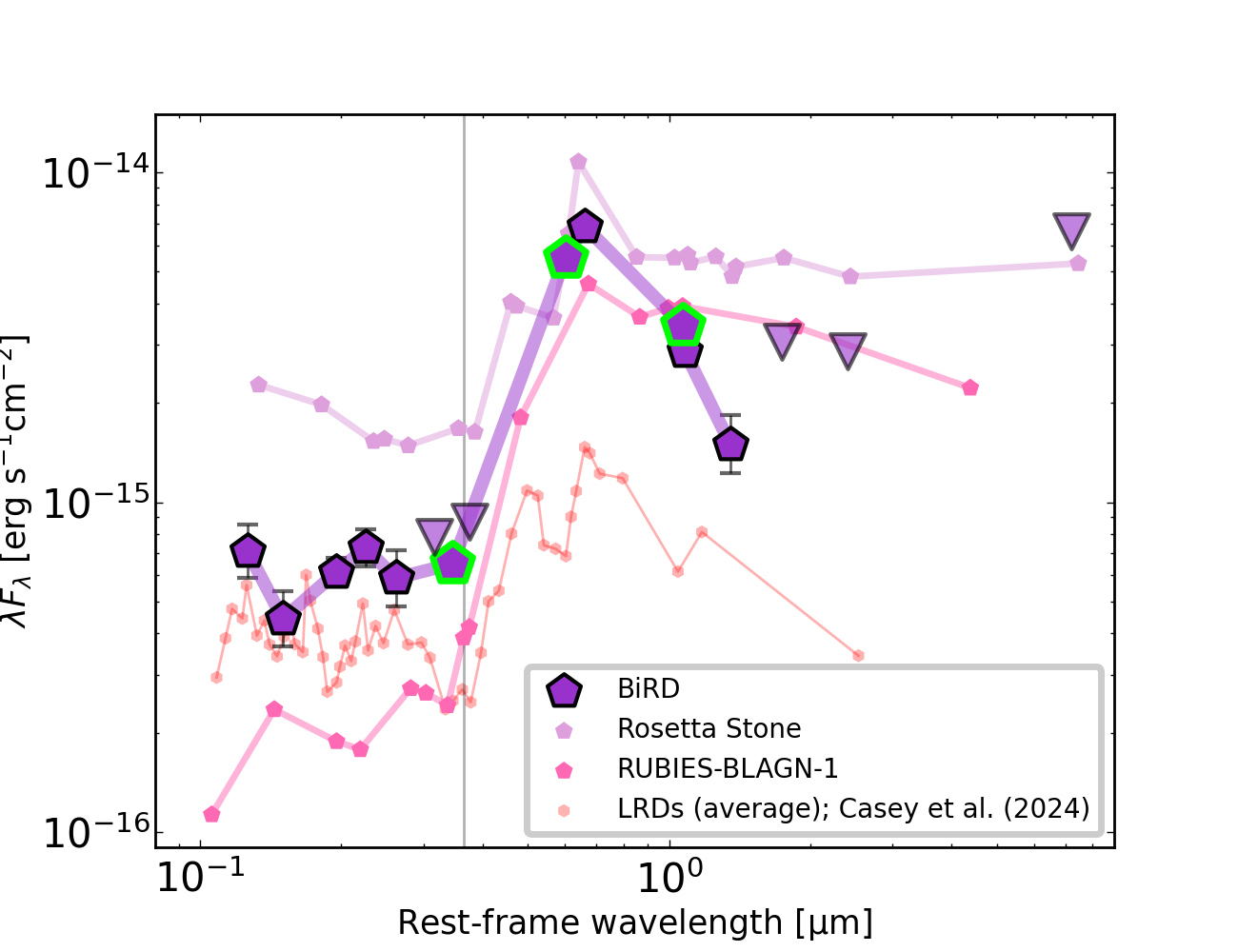}\\
\end{center}
\caption{Broad band, rest-frame SED of BiRD (purple symbols) compared to that of the Rosetta Stone (plum; \citealt{juodzbalis24}), RUBIES-BLAGN-1 (pink; \citealt{wang25}) and to the median SED of LRDs at $3 \lesssim z \lesssim 9$ (red; \citealt{casey24}). The green contour highlights the JWST data-points at $1.15$, $2.0$, and $3.56\ \rm \mu m$, while the inverted triangles represent flux upper limits. The photometric errors for BiRD are also reported and, for some data-points, are smaller than the symbol size. The position of the Balmer break is marked by the vertical line at 3645 \AA.}
\label{fig_sed}
\end{figure}

\subsection{Black hole mass}
\label{sub:bh}
We derive the black hole mass of BiRD using two independent approaches, which both start from the \Pag\ broad line. In the first one, we use Eq. (2) from \citet{landt13}, which employs the \Pag\ FWHM and the $1\ \mu \rm m$ rest-frame continuum to estimate the black hole mass
{\small
\begin{equation}
    \log \left( \frac{M_{\rm BH}}{\rm M_{\odot}} \right) = 0.88 \left[2 \log \left(\frac{FWHM(\rm Pa \gamma)}{\rm km\ s^{-1}}\right) + 0.5 \log \left( \frac{\nu L_{1 \mu m}}{\rm erg\ s^{-1}} \right) \right] -17.39.
\label{eq:landt13}
\end{equation}
}
We use the $FWHM$ of the broad \Pag\ component determined in our fit. Unfortunately, the $1\ \mu \rm m$ rest-frame continuum is outside the range of our spectrum. However, we note that the fitted continuum appears quite flat (Fig.~\ref{fig_fit} and Fig.~\ref{fig_2dspec}), therefore, we use the continuum underlying \Pag\ (i.e., $1.09\ \mu \rm m$ rest-frame) as a surrogate. Using Eq.~\ref{eq:landt13}, we obtain a black hole mass of 
\begin{equation*}
    M_{\rm BH}(\rm Pa\gamma) = (7 \pm 2) \times 10^7\ \rm M_{\odot}
\end{equation*}
While the formal error on the BH mass of the estimate above is around the 30\%, we note that the calibration of \citet{landt13} is characterized by an intrinsic scatter of $\sim 1\ \rm dex$.

An alternative approach to derive the black hole mass is based on using the \Pag\ broad component to infer the properties of the \Ha\ broad line. The latter is more often used as a black hole mass tracer in low-redshift active galaxies. Using the relation of \citet{reines&volonteri15}, we can estimate the black hole mass as
{\small
\begin{equation}
    \log \left( \frac{M_{\rm BH}}{\rm M_{\odot}} \right) = 6.60 + 0.47 \log \left( \frac{L(\rm H \alpha)}{10^{42} \rm erg\ s^{-1}} \right) + 2.06 \log \left( \frac{FWHM(\rm H \alpha)}{1000\ \rm km\ s^{-1}} \right)
\label{eq:rv15}
\end{equation}
}
Adopting a different calibration, i.e., \citet{greene&ho05}, systematically lowers the black hole masses by a factor of $\sim 2$. 

We based our argument on the analogy between BiRD and Rosetta Stone. As previously mentioned (see Sect.~\ref{sub:xrp}), the hypothesis of case B recombination seems to work well for the Rosetta Stone. The broad \Pag\ in the Rosetta Stone appears not to be significantly dust-attenuated, as the observed broad line ratio of \Pag\ and \Pab\ is $\sim 0.525$, very close to the intrinsic one ($\sim 0.55$; \citealt{osterbrock89}).  
Thus, based on the intrinsic \Ha\ and \Pag\ flux ratio $\sim 31.7$ \citep{osterbrock89}, we can infer the unextincted \Ha\ broad line luminosity $\sim 7 \times 10^{43}\ \rm erg\ s^{-1}$ from the observed broad \Pag\ line in the Rosetta Stone. Substituting in Eq.~\ref{eq:rv15}, we find for the Rosetta Stone a black hole mass $\log{(M_{\rm BH}/\rm M_{\odot})} \sim 8.6$. This estimate is nicely consistent with the value obtained from the measured \Ha\ luminosity, corrected for dust extinction based on the observed \Ha /\Hb\ broad line ratio, i.e., $\log{(M_{\rm BH}/\rm M_{\odot})} = 8.67 \pm 0.30$, which becomes $\log{(M_{\rm BH}/\rm M_{\odot})} = 8.47 \pm 0.30$, if the extinction correction is based on the narrow lines \citep{juodzbalis24}.

We thus apply the same argument to BiRD. The inferred broad \Ha\ luminosity from the observed \Pag\ is in this case $L(\rm H \alpha_{\rm est}) = 2.6^{+0.3}_{-0.4} \times 10^{43}\ \rm erg\ s^{-1}$.

Another ingredient that goes in Eq.~\ref{eq:rv15} to estimate the black hole mass is the $FWHM$ of the \Ha\ line. We note that in the Rosetta Stone the ratio between the $FWHM$ of broad \Ha\ and \Pag\ lines is $\sim 1.57$. We assume that the same ratio holds for BiRD. On the other hand, \citet{wang25} found a lower ratio $FWHM(\rm H \alpha)/FWHM(\rm Pa \gamma) \sim 1.43$ in RUBIES-BLAGN-1. We account for this difference when computing the errors on the black hole mass. By applying Eq.~\ref{eq:rv15}, we find
\begin{equation*}
    M_{\rm BH}(\rm H\alpha_{est}) = 2.4^{+1.0}_{-1.3} \times 10^8\ \rm M_{\odot} 
\end{equation*}
We note that the calibration of \citet{reines&volonteri15} has an intrinsic scatter $\sim 0.5\ \rm dex$.
This estimate of the black hole mass of BiRD is fully consistent with the value that we obtained adopting the relation of \citet{landt13}, taking into account the $1\ \rm dex$ scatter of the latter relation. Using the calibration of \citet{greene&ho05} would result in a decrease of the black hole mass by a factor of $\sim 2$, lowering the difference between the two estimates.

In Sect.~\ref{sub:bhmassfunc}, where we discuss the active black hole mass function (BHMF) of LRDs, we use the black hole mass $M_{\rm BH}(\rm H\alpha_{est})$ of BiRD derived from \Ha\ using the calibration of \citet{reines&volonteri15} to ease the comparison with other works. 
\begin{figure}
\begin{center}
\includegraphics[width=0.5\textwidth]{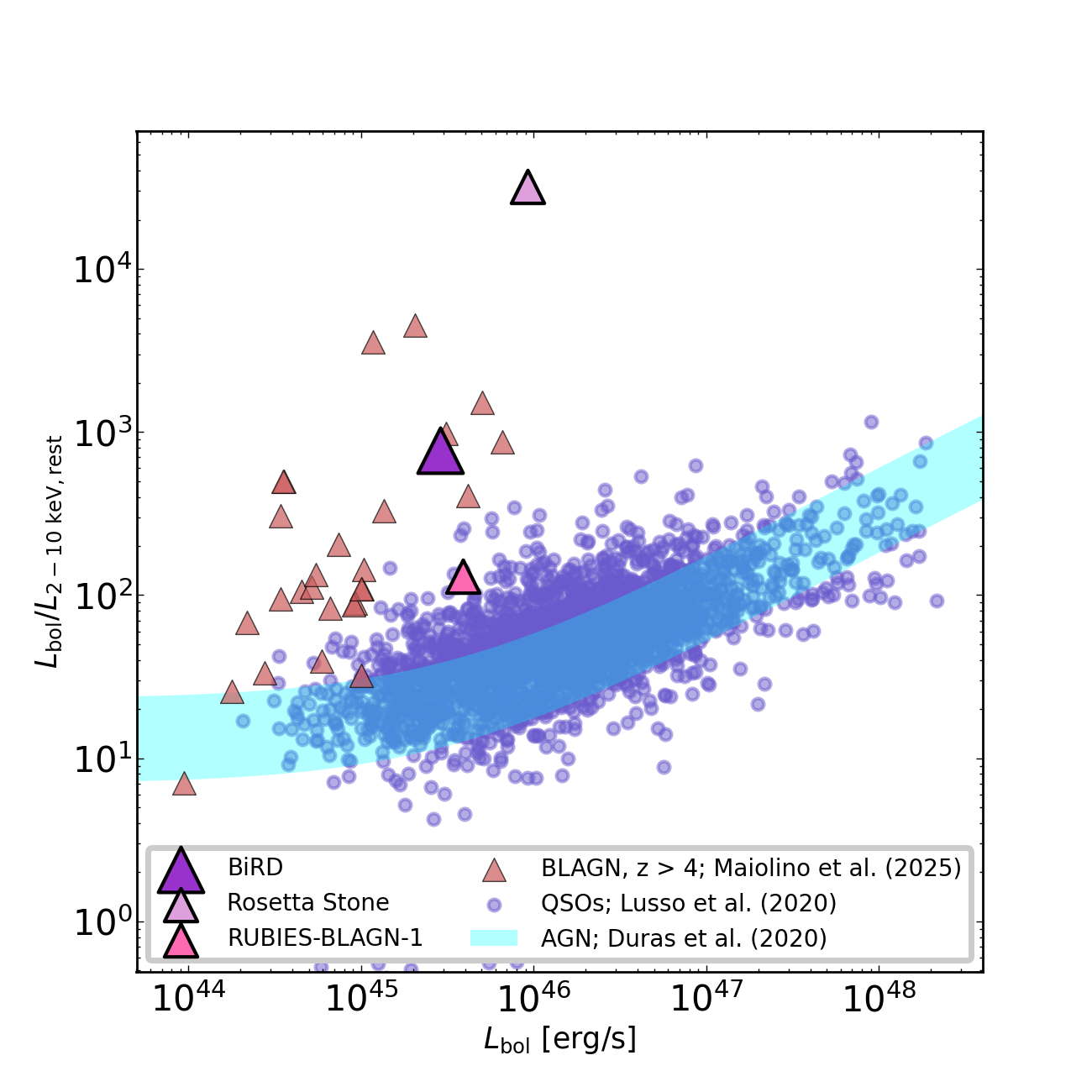}\\
\end{center}
\caption{X-ray (2-10 keV) bolometric correction vs bolometric luminosity for the standard population of broad line quasars (slate blue filled circles; \citealt{lusso20}) and the new population of X-ray silent, broad line AGN discovered by JWST at $z>4$ \citep{maiolino25}. The latter is shown with red triangles, which represent lower limits to the bolometric correction. The cyan area shows the relation found by \citet{duras20} for broad line AGN with its dispersion. Similarly to the new JWST population, the BiRD is undetected in the X-rays (purple triangle), and its X-ray bolometric correction is at least 30 times larger than expected in standard quasars. The lower limits to the bolometric correction for the Rosetta Stone (in plum) and RUBIES-BLAGN-1 (pink triangle) are also shown for comparison.}
\label{fig_xray}
\end{figure}

\subsection{\Hei\ absorption}
\label{sub:abs}
We estimate the column density of \Hei\ atoms that generate the absorption in the BiRD spectrum (Fig.~\ref{fig_fit}). As seen in Sect.~\ref{sub:fit}, we start considering a scenario in which the absorber acts on both the continuum radiation and the BLR (see also \citealt{juodzbalis24}). The absorption of the bound-bound transition $2^3 S - 2^3 P$ of \Hei\ depends on the number density along the line of sight (column density). The column density $N$(\Hei) can be estimated using the apparent optical depth method \citep{savagesembach91}. Given $N(v)$, the column density per unit velocity, we integrated Eq. (8) of \citet{savagesembach91} and solved for $N = \int N(v) dv$:
\begin{equation}
    N({\rm He\ I}) = \int N(v) dv = 3.768 \times 10^{14} (f\lambda_{\rm rest})^{-1} \int \tau(v) dv\ \rm [cm^{-2}]
    \label{eq:densform}
\end{equation}
where $f$ and $\lambda_{\rm rest}$ are the oscillator strength and the rest-frame wavelength of the \Hei\ line expressed in \AA\ ($10830\ \AA$), respectively, and $\tau(v)$ is the optical depth as a function of velocity. We consider the oscillator strength for the $2^3 P$ to $2^3 S$ transition from \citet{drake06}, $f = 0.5390861$ (see their Table 11.11). We integrate Eq.~\ref{eq:densform} and find 
\begin{equation}
    N(\rm He\ I) = 1.2^{+1.5}_{-0.4} \times 10^{14}\ \rm cm^{-2}
    \label{eq:densval}
\end{equation}
 which corresponds to the column density of \Hei\ atoms in the $2^3 S$ level. We note that the absorption is blueshifted by $\Delta v = -830_{-148}^{+131}$ \kms\ (see Table~\ref{tab:abs}). This suggests that the absorber is outflowing from the nuclear region.

We also evaluate how a different location of the absorber affects the derived \Hei\ column density. As we discuss in Sect.~\ref{sub:fit}, our data cannot constrain the location of the absorber and it is thus necessary to consider other scenarios. We repeat the fit in Sect.~\ref{sub:fit} by applying the absorption term also to the narrow \Hei\ component, i.e., assuming that the absorber is located outside of the NLR (see also \citealt{wang25}, who considered a similar hypothesis). In this case, we obtain a lower value of the column density, $N(\rm He\ I) = 4.7^{+1.0}_{-0.8} \times 10^{13}\ \rm cm^{-2}$. This difference is expected as the absorber is now acting on the entire emitted flux (from the continuum, BLR and NLR) and thus has a lower optical depth. 
Also in this case the absorption is blueshifted, with $\Delta v = -949_{-147}^{+148}$ \kms , meaning that the gas is outflowing. We note that the absorber location does not affect the blueshifted velocity, which is consistent within the errors in the two different assumptions. 

\subsection{Space density of LRDs at cosmic noon}
\label{sub:spaceden}

The discovery of the BiRD adds to that of the Rosetta Stone and RUBIES-BLAGN-1, leading us to attempt a first measurement of the space density of LRDs around cosmic noon based on JWST data. 
The BiRD has been selected as a bright outlier in the color–magnitude diagram of point-like sources in the J1030 field (see Fig.~\ref{fig_dplot}), as it was both falling above the stellar sequence and sufficiently distant from the locus occupied by fainter, high-z LRDs. The selection box can be described as follows: $F200W-F356W >-0.4$, $F356W < 23.0$. Both the Rosetta Stone and RUBIES-BLAGN1 satisfy these selection criteria (Fig.~\ref{fig_dplot}). 

Based on the observed SED of the BiRD, we estimated the color-magnitude track that an LRD would follow when moved at different redshifts: for $z<z_{\rm min}\sim2$, LRDs would mix with stars and would not be selected by their $F200W-F356W$ color. Also, in addition to the applied limiting magnitude ($F356W < 23$) to select relatively luminous objects, we applied another color cut at $F200W-F356W < 2$ to consider only LRDs at $z\lesssim3.5$ (see Fig.~\ref{fig_dplot}). This redshift limit is also consistent with the SEDs of the other two LRDs.

Since we are dealing with objects that are well above the flux limit of the survey fields in which they were discovered, we used the simple $1/V_{\rm max}$ method \citep{schmidt68} to estimate their number density:

\begin{equation}
\label{eq:phi}
\Phi(L) = \frac{1}{\Delta \log L} \sum_{\rm i} \frac{w_{\rm i}}{V_{\rm max,i}},
\end{equation}

where $\Delta \log L$ is the width of the luminosity bin, $w_{\rm i}$ are weights that account for incompleteness effects, and $V_{\rm max,i}$ is the maximum volume in which an object i would have been detected, i.e.

\begin{equation}
\label{eq:vmax}
V_{\rm max,i} = f_\Omega \int_{z_{\rm min,i}}^{z_{\rm max,i}} \frac{dV_{\rm c}}{dz} dz ,
\end{equation}

where $dV_{\rm c}/dz$ is the comoving volume element and $f_\Omega$ is the fraction of the solid angle of the sky covered by the observations.
The confidence intervals in $\Phi(L)$ are computed following the standard approximations of \citet{gehrels86} for low count statistics. 

We adopt $z_{\rm min} = 2.0$ and $z_{\rm max}=3.5$ for all three objects. The correspondence between the adopted color selection criteria and the effective redshift boundaries clearly depends on the chosen SED track, but we note that statistical uncertainties greatly exceed those produced by the choice of the exact redshift limits. 

Rather, an important source of uncertainty is related to the completeness of our sample. The J1030 (EIGER), JADES, and RUBIES survey fields are covered by deep JWST imaging of comparable depth, and any of the three considered objects would have been detected in any of the considered survey fields. However, other sources in the JADES and RUBIES fields might satisfy our selection criteria, hence $w_{\rm i} \geq 1$ for both Rosetta Stone and RUBIES-BLAGN-1. In our estimate, we will conservatively assume $w_{\rm i}=1$, and treat our measurement of the space density of $z\sim2.5$ LRDs as a lower limit.

The total solid angle considered is $\Omega_{\rm tot} = \Omega_{\rm J1030} + \Omega_{\rm JADES} + \Omega_{\rm RUBIES} = 302$ arcmin$^2$ , with $\Omega_{\rm J1030}, \Omega_{\rm JADES}, \Omega_{\rm RUBIES}$ = 27, 125, and 150 arcmin$^2$, respectively, and corresponds to $f_\Omega =2.03\times 10^{-6}$. 

To compute $\Phi(L)$, we considered the bolometric luminosities of the three objects derived from their broad Balmer or Paschen lines and used a luminosity bin of 0.5 dex as it spans the full luminosity range covered by the three targets (3, 3.9, 9.3$\times 10^{45}\ \ergs$ for the BiRD, RUBIES-BLAGN-1, and the Rosetta Stone, respectively). 
This returns $\Phi(L) = 4.0^{+4.0}_{-2.4} \times 10^{-6}$ Mpc$^{-3}$ dex$^{-1}$ at $L_{\rm bol,mean}= 5.4 \times 10^{45}\ \ergs$ and $z_{\rm mean}\sim2.5$ (see Fig.~\ref{fig_phi}).

Admittedly, the centering of the 0.5-dex luminosity bin has been chosen to include all three sources, and is therefore arbitrary. If a random positioning of 0.5-dex-wide bins was adopted, the three sources would split into two adjacent bins, returning, on average, space densities $\sim 2$ times lower than the measured value. Given the very small source statistics, all measurements would be nonetheless consistent with each other within the errors, and therefore we adopt the quoted value of $\Phi(L) = 4.0^{+4.0}_{-2.4} \times 10^{-6}$ Mpc$^{-3}$ dex$^{-1}$ as our reference measurement.

\section{Discussion}
\label{sec:disc}
\subsection{The absorber}
\label{sub:absorberdisc}

In Sect.~\ref{sub:abs} we estimated the column density of \Hei\ that produces the observed absorption in the BiRD spectrum. As mentioned before, we note that, based on our data we cannot constrain the location of the absorber. \citet{juodzbalis24} also find in the Rosetta Stone blueshifted absorption feature in the \Hei\ line profile ($\Delta v = -506 \pm 7$ \kms\ and $FWHM = 714 \pm 14$ \kms), and in the broad Balmer lines. Due to its resonant nature, the \Hei\ transition is much more easy to see in absorption compared to Balmer lines, which require higher gas and column density. 

For the Rosetta Stone, two scenarios are discussed, the first one considering a single absorber producing both absorptions and the second considering two separate absorbers, with the highest density one being responsible for Balmer absorption. In both scenarios, based on CLOUDY modeling, they estimate the hydrogen density $n_{ \rm H} \gtrsim 10^8\ \rm cm^{-3}$, which favors a BLR origin for the Balmer absorbing clouds. Assuming the same origin also for the \Hei\ absorbing clouds, \citet{juodzbalis24} find in the Rosetta Stone a \Hei\ column density in the $2^3S$ level of $N(\rm He\ I) = 3.63^{+0.08}_{-0.04} \times 10^{14}\ \rm cm^{-2}$. A BLR origin for the absorbing gas was adopted also by \citet{deugenio25, i&m25, ji25}. In the same scenario, we derive for BiRD $N(\rm He\ I) = 1.2^{+1.5}_{-0.4} \times 10^{14}\ \rm cm^{-2}$, which is a factor $\sim 3$ lower than the value seen in the Rosetta Stone.

Blueshifted \Hei\ absorption was also found by \citet{wang25} in RUBIES-BLAGN-1. The absorption is blueshifted by $-863_{-113}^{+121}$ \kms and a $FWHM$ of $569 \pm 23$ \kms . The blueshift seen in RUBIES-BLAGN-1 is very similar to that observed in BiRD ($-863_{-113}^{+121}$ \kms), while our $FWHM$ is lower ($310_{-68}^{+77}$ \kms). Regarding the column density, \citet{wang25} assumes that the absorber acts on the entire emitted radiation (continuum, BLR and NLR). They find a \Hei\ column density $N(\rm He\ I) = (5.9 \pm 0.3) \times 10^{13}\ \rm cm^{-2}$. This value is consistent with the value that we find for the BiRD under the same hypothesis, i.e., in the case of an absorber located outside of the NLR ($N(\rm He\ I) = 4.7^{+1.0}_{-0.8} \times 10^{13}\ \rm cm^{-2}$). 

Similar \Hei\ absorption was previously detected in active galaxies and quasars at low redshift \citep{hutchings02,landt19,pan19}. Several works about JWST-discovered broad line AGN report Balmer absorption, with a detection rate that is much higher compared to low-redshift AGN \citep{juodzbalis24, kocevski25, lin24, maiolino24, matthee24}. Unfortunately, due to the limited wavelength coverage of our data, we cannot verify if Balmer absorptions also affect the spectrum of BiRD, and we cannot estimate the hydrogen density. 

The similarity in the \Hei\ absorption shown by BiRD, Rosetta Stone, and RUBIES-BLAGN-1 suggests that this feature could be common in LRDs at cosmic noon. Thus, observing the \Hei\ line in more objects at $z\sim 2-3$ will be crucial for a better understanding of the properties of the absorbing gas. In principle, it is possible that the \Hei\ absorption also affects the spectra of high-redshift ($z > 4$) LRDs, as it has been seen for Balmer absorption. Unfortunately, this feature can be seen with NIRSpec gratings and NIRCam grisms only up to $z \lesssim 4$, which may explain why the detections were reported only at $z \sim 2-3$. 

\subsection{SED shape, dust content, and optical extinction}
\label{sub:sed-discuss}

As shown in Fig.~\ref{fig_sed}, the SED of the BiRD is very similar to that of the Rosetta Stone and of RUBIES-BLAGN-1, as well as to the median SED of LRDs at higher redshifts \citep{casey24}, at least below 1 $\mu$m rest-frame. Above that wavelength, the BiRD SED has a marked downturn that is not observed in the typical SED of LRDs. Since the BiRD is detected only up $\lambda_{\rm rest}\sim 1.3\ \mu$m (IRAC-CH2), it is unclear whether and how strongly this downturn would continue at longer wavelengths. In any case, the current upper limits at $\sim 1.7 - 6.3\ \mu \rm m$ in the rest frame already indicate that the BiRD's emission at those wavelengths cannot be as strong as that of the Rosetta Stone or RUBIES-BLAGN-1.

In RUBIES-BLAGN-1, the relatively flat - or even declining - SED (in terms of $\lambda f_\lambda$) at rest-frame wavelengths $\lambda > 1\ \rm \mu m$ has been interpreted by \citet{wang25} as evidence for the absence of emission from hot dust at temperatures around $1000-1500\ \rm K$. This hot dust, typically located in the parsec-scale circumnuclear torus, is usually responsible for the rising AGN SED at $\lambda_{rest} > 1\ \rm \mu m$ \citep{richards06,trefoloni24}.
The downturn observed in the BiRD SED above that wavelength would argue even more strongly against the presence of a hot dusty torus. 

In general, however, observations of LRDs mid-IR SEDs do not rule out the presence of circumnuclear dusty tori with different physical or geometrical properties, such as clumpy tori where the self-shielding of individual clouds leads to colder mean temperatures \citep{nenkova08}, or tori with dust grain composition skewed towards large grains and extended on tens of pc scales \citep{li25}. In any case, the average SED of LRDs points towards an overall low dust content in these systems. 

Based on the compact sizes of the sources and assuming that dust reddening is responsible for the observed SED shape just above the Balmer break, \citet{casey24} estimated an average dust mass in high-z LRDs of $M_{\rm d} \propto R_{\rm d}^2 N_{\rm d} \sim10^4$\Msun, where $R_d$ and $N_d$ are the effective radius and column density of the dust distribution, respectively, and $N_{\rm d}\propto A_{\rm V}$. Such dust‑poor systems are significantly rare at any epoch. In the nearby Universe they are confined to a handful of extremely metal‑poor blue‑compact dwarfs \citep{fisher14, calura25}.
At earlier times the census is even sparser: recent ALMA studies reveal only a few high‑redshift galaxies with similarly weak dust emission, implying gas‑to‑dust ratios $\sim 300-1000$ (e.g., \citealt{spilker25}). However, the limitation of the approach of \citet{casey25} is that the average dust column density is measured along the line of sight, which is sensitive to the geometric distribution of dust and is likely biased towards low column densities.

Anyhow, even if the estimated dust masses may be regarded as lower limits, all the observational evidences point towards a low content of dust at any temperature in LRDs.
All the attempts to detect the mm/sub-mm continuum in these systems using ALMA - including RUBIES-BLAGN-1 \citep{akins25,setton25} - have been unsuccessful, placing upper limits on the total dust mass at temperatures of a few tens of K of $M_{\rm d} \lesssim 10^8\ \rm M_\odot$ from individual observations, and of $M_{\rm d} \lesssim 10^7\ \rm M_\odot$ from a stacked sample of 60 LRDs \citep{casey25}.

Dust extinction in LRDs is generally measured through SED fitting, assuming that the observed red continuum results from dust attenuation of an intrinsically blue spectrum produced by stars or an AGN. By using this technique, \citet{kokorev24} report for dust extinction an average $A_{\rm V} = 1.6$ in their sample, while \citet{akins24} find values that can be $\gtrsim 3\ \rm mag$ in several objects. The differences in the $A_{\rm V}$ values are possibly due to different selections of LRDs. High $A_{\rm V}$ values hardly reconcile with the results on individual objects reported by \citet{chen25}, who find upper limits to the optical extinction values of $A_{\rm V} \lesssim 1.0-1.5\ \rm mag$ by modeling the dust reprocessing with various extinction curves and incident spectra.

An alternative way to derive $A_{\rm V}$ is via Balmer decrement. Assuming case B recombination and using the ratio of narrow lines, the values of $A_{\rm V}$ are in the range of $A_{\rm V}\sim 0.5-1.4$ for individual LRDs as RUBIES-BLAGN-1\footnote{For RUBIES-BLAGN-1, we estimate $A_{\rm V}$ from the \Ha\ and \Pag\ flux reported in Table~4 of \citet{wang25}, assuming unextincted \Pag\ flux and using either Small Magellanic Cloud \citep{gordon03} or Calzetti \citep{calzetti00} extinction law, for both narrow and broad lines.} \citep{wang25}, the Rosetta Stone \citep{juodzbalis24}, and BlackTHUNDER \citep{deugenio25}. Using broad lines increases these values to $A_{\rm V} \lesssim 2.3-3.1$\footnote{We note that \citet{deugenio25} assume for broad lines an intrinsic \Ha /\Hb\ ratio of 3.1 \citep{dong08}, which is different from case B. However, this assumption does not significantly impact the value of $A_{\rm V}$}.

Based on the reasoning above, we caution that incorrectly high $A_{\rm V}$ values from Balmer lines, possibly due to deviations from case B recombination, may lead to an overestimate of the intrinsic \Ha\ luminosity, and thus to a corresponding overestimate in the inferred bolometric luminosity and black hole mass.

\subsection{X-ray weakness, obscuration and black hole accretion}
\label{sub:edd}

As reported in Section~\ref{sub:xrp}, the BiRD is not detected in the deep Chandra image of the field. This places an upper limit to its $2-10\ \rm  keV$ rest-frame luminosity of $L_{\rm 2-10\ keV,rest}<3.7\times 10^{42}\ \ergs$, which is at least $30$ times lower than what would be expected for a standard broad line AGN of equal bolometric luminosity (\citealt{duras20}; see Fig.~\ref{fig_xray}). Similarly to what has been observed in high-z LRDs and in the BLAGN population discovered by JWST, the BiRD is therefore surprisingly weak in the X-rays, and the reasons for this widespread X-ray weakness in LRDs are still under scrutiny. 

If the BiRD had standard, type-1 AGN bolometric corrections, to depress its intrinsic $2-10\ \rm keV$ luminosity by a factor of $\sim 30$, absorption by gas column densities of at least $10^{24}\ \rm cm^{-2}$, that is Compton-thick, is needed \citep{gilli07,lanzuisi18}.\footnote{This estimate assumes absorption by gas with solar abundances. Sub-solar abundances, often observed in the ISM of JWST high-$z$ galaxies and AGN \citep{curti24}, would lead to even larger densities.} Heavy, Compton-thick absorption by dust-poor or dust-free gas such as that in the BLR clouds or their immediate surroundings has indeed been proposed to explain the widespread X-ray weakness of LRDs \citep{maiolino25}, provided that the BLR covering factor is significantly larger than in standard AGN. The large equivalent width of the broad \Ha\ line commonly observed in JWST BLAGN, as well as the frequent presence of an absorption component in the Balmer lines and of a prominent Balmer break in those systems featuring high-quality spectra, may corroborate this interpretation \citep{maiolino25,juodzbalis25}.

In addition to heavy nuclear absorption, it is often hypothesized that JWST BLAGN are X-ray weak because they accrete at super-Eddington rates \citep{madau24, pacucci24, king25, madau25}.
At high accretion rates, geometrically thick disks may form around black holes, leading to a highly anisotropic emission. At large viewing angles from the polar axis, the innermost, hottest regions of the accretion flow (inner disk and corona) are shadowed by the outer cooler regions of the thick disk, leading to a significant luminosity decrease, especially in the X-rays \citep{madau25}.

Furthermore, in the above geometry, the UV photon density is such to Compton-cool the electrons in the corona much more efficiently than in standard thin disk plus corona models, making the emerging X-ray spectrum very soft ($\Gamma\sim3-4$; \citealt{pacucci24,madau24}). These two effects lead these models to predict $2-10\ \rm keV$ bolometric corrections of $k_{2-10\ \rm keV}>100-1000$ \citep{madau24,pacucci24}, similar to what is observed. However, we note that the average Eddington ratio measured for JWST BLAGN is $\lambda_{\rm E} \sim 0.1$ \citep{juodzbalis25} and the X-ray weakness persists also for "dormant" AGN discovered by JWST at high-$z$, accreting at only 1\% of the Eddington limit \citep{juodzbalis24b}, suggesting that super-Eddington accretion may not be the only explanation for the X-ray weakness for the whole population.

When accounting for the large uncertainties in its estimated bolometric luminosity and black hole mass, we infer for the BiRD an Eddington ratio of $\lambda_{\rm E} = 0.25^{+0.25}_{-0.20}$. Similar values are measured for both the Rosetta Stone and RUBIES-BLAGN-1.
In the thick-disk super-Eddington model by \citet{madau25}, such low values of the apparent Eddington ratio can be obtained only at inclinations larger than $80$ deg, meaning that most JWST AGN should be observed almost edge-on. This can hardly be reconciled with random disk orientations: at 57 degrees - the average viewing angle for random disk inclinations - one would expect in this model $\lambda_{\rm E} \sim 3$.

However, in other studies of super-Eddington accretion flows, the decrease of the apparent luminosity with the thick disk inclination is much more rapid, owing to the very narrow opening the funnel formed by the accretion flow \citep{ohsuga05}, and this tension might be alleviated. We also note that super-Eddington accretion may still be tenable if the BH masses were overestimated by one or two orders of magnitude. This has been in fact proposed by recent works suggesting that either Balmer scattering \citep{naidu25} or electron scattering \citep{rusakov25} from a layer of very dense and ionized gas around the BLR may broaden the Balmer lines far above their kinematic width.

While these scenarios are intriguing, they await confirmation as they seem in contrast with the finding that the same profiles are seen in local AGN, whose BH masses area measured directly, and may even be in contrast with some features observed in the Paschen and Balmer lines of JWST BLAGN (see \citealt{juodzbalis25} for a full discussion). Moreover, for obscured AGN candidates at $z\sim 6$ in the SHELLQs survey \citep{matsuoka25}, which exhibit narrow Ly$\alpha$ lines and broad H$\alpha$ components that are significantly weaker than their narrow counterparts compared to local quasars, it has recently been suggested that the BLR is partially hidden \citep{iwasawa25}. If this is the case also for LRDs, then their BH masses may have been instead underestimated, making these objects even more puzzling.  

\subsection{Space density and evolution of LRDs}
\label{sub:spacedenevol}

\begin{figure*}
\begin{center}
\includegraphics[width=1\textwidth]{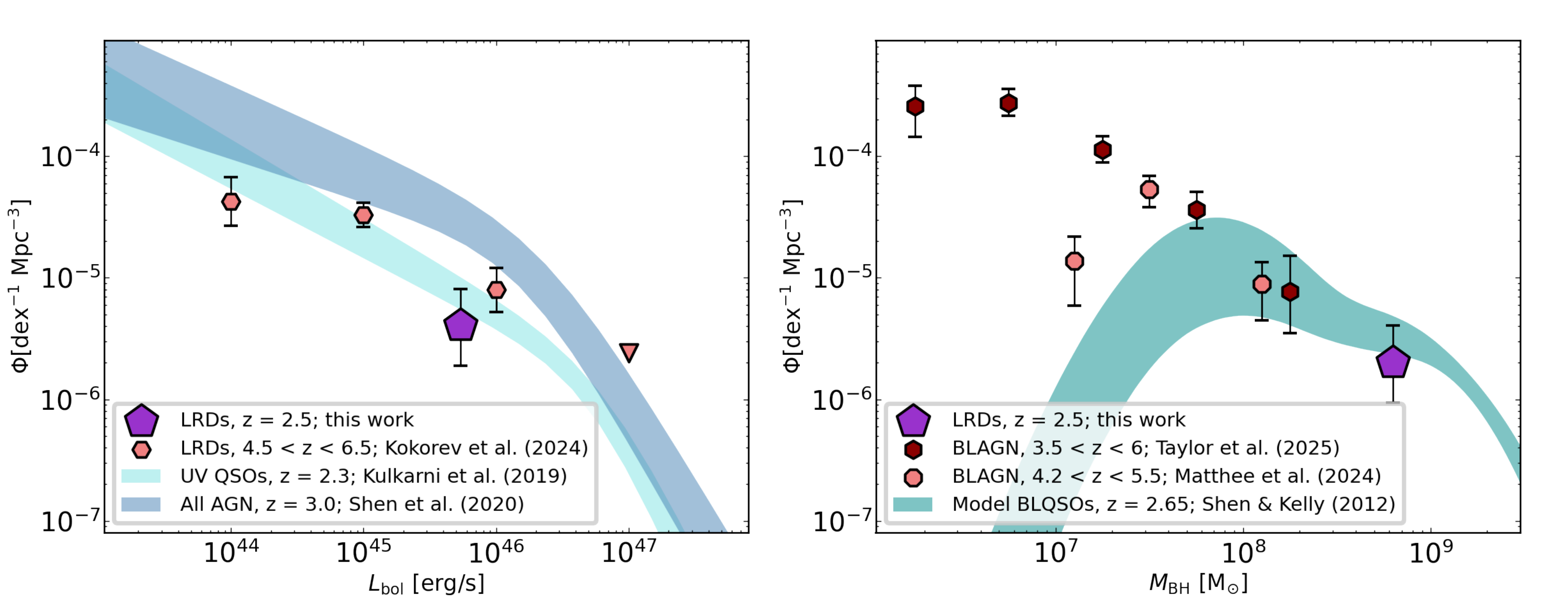}\\
\end{center}
\caption{\textit{Left}: Space density of LRDs at $L_{\rm bol,mean}= 5.4 \times 10^{45}\ \ergs$ and $z_{\rm mean}\sim2.5$ from our work (purple pentagon) compared with the bolometric luminosity function derived at similar redshifts for the entire AGN population (steel blue; \citealt{shen20}) and for blue, UV-selected, unobscured AGN (pale turquoise; \citealt{kulkarni19}). The shaded areas correspond to the 16th and 84th percentiles of the bolometric LFs. At cosmic noon, the measured abundance of LRDs appears to be a factor $2-3$ lower than that of unobscured AGN. We also compare our estimate with the bolometric luminosity function of LRDs at $4.5 < z < 6.5$ from \citet{kokorev24} (in red). \textit{Right}: Space density of LRDs at $z_{\rm mean}\sim2.5$ and $M_{\rm BH} \sim 6.3 \times 10^8\ \rm M_{\odot}$ (purple) compared with the predicted active BHMF from broad line quasars at similar redshift (dark cyan; \citealt{shenkelly12}) and the active BHMF of LRDs at higher redshifts (red symbols, \citealt{matthee24, taylor24}). The shaded area corresponds to the 16th and 84th percentiles of the model active BHMF value.} 
\label{fig_phi}
\end{figure*}

In Fig.~\ref{fig_phi} (left panel) we compare our measured space density of LRDs at $z\sim2.5$ with the bolometric luminosity function of all AGN \citep{shen20} and UV-selected QSOs \citep{kulkarni19} at similar redshifts. 
\citet{shen20} considered AGN samples selected at different wavelengths, from the IR to the X-rays and, based on a detailed SED modeling, derived bolometric and extinction corrections for all
objects in their samples. Their measurements are therefore supposed to include both populations of obscured and unobscured AGN, providing an estimate of the bolometric luminosity function of the whole AGN population and of its evolution up to $z\sim 7$.
Based on a large sample of quasars selected by their blue UV/optical colors, \citet{kulkarni19} estimated the luminosity function at 1450 \AA\ rest-frame of unobscured AGN up to $z\sim 7.5$. To convert their measurements from UV to bolometric AGN luminosities, we used a bolometric correction of $L_{\rm bol}/L_{1450} = 4.0$, consistent e.g., with the estimates of \citet{richards06} and \citet{runnoe12}.

As shown in Fig.~\ref{fig_phi}, the density of LRDs at $z\sim2.5$ is only a factor of $\sim 2-3$ lower than that of standard UV-selected quasars with similar bolometric luminosity, and consistent with it within the errors. As already mentioned in Sect.~\ref{sub:spaceden}, the measured abundance of LRDs could even be treated as a lower limit. In fact, other LRDs at $z\sim 2-3$ awaiting for spectroscopic confirmation might be present in the JADES and/or RUBIES fields, each covering a factor of $\sim 5$ larger area than the EIGER observation of J1030.

However, we caution that it remains unclear whether the presence of similar objects has already been ruled out in other JWST fields, which would instead bias our measurement to higher values. As the analysis of JWST data in well-studied survey fields progresses, the number of spectroscopically confirmed LRDs at $z\sim2-3$ is steadily increasing \citep{hviding25, zhuang25}. We therefore consider it reasonable to assume that LRDs at $z\sim2-3$ in other JWST deep survey fields have yet to be discovered primarily because of the current incompleteness of spectroscopic analysis and follow-up, and that the three fields analyzed in this study offer a representative sampling of the LRD population at these redshifts. 

To date, the vast majority of LRDs, including those confirmed spectroscopically, have been discovered at $z>4$.
Their abundance in those epochs has been found to largely exceed that of standard AGN, even by up to one order of magnitude \citep{kocevski25}. However, a fair comparison between the luminosity function of LRDs and standard AGN is complicated by the still unknown physical properties of the former, which makes any estimate of their intrinsic bolometric luminosity highly uncertain. For example, it is unclear whether their observed rest-UV luminosities can be reliably used to infer their intrinsic bolometric luminosities. The measured rest-UV luminosities of LRDs are in fact significantly lower than the expected {\it intrinsic} value derived by fitting their optical/NIR SEDs with standard AGN templates and once correcting for extinction, yet higher (by up to two dex) than what the best fit extinction would produce \citep{kokorev24, matthee24}. 
The same uncertainties apply to the UV luminosity function recently derived for the LRD candidates at all redshifts selected through Euclid photometry \citep{bisigello25}.

Here we consider H$\alpha$ emission as a more reliable proxy of the LRD bolometric luminosity, as it is less prone to extinction effects than UV emission and it is directly related to the total, intrinsic UV photon production. However, we note that at high neutral hydrogen column densities, such as those proposed by \citet{i&m25} for LRDs, a fraction of the \Ha\ photons can be absorbed by electrons in the $n=2$ energy level. This would depress the observed \Ha\ line luminosity, thus further increasing the uncertainty related to the bolometric luminosity and black hole mass of LRDs from the observed \Ha .

Our measurement of the bolometric luminosity function of LRDs at $z\sim 2.5$ can be compared with that measured by \citet{kokorev24} at $z>4$ (which, however, is derived from the SED fitting of candidate LRDs with photometric redshift only). At $L_{\rm bol}= 5.4 \times 10^{45}\  \ergs$, the abundance of LRDs increases from $z\sim 2.5$ to $z=5.5$ by a factor of 4. At $z=5.5$ their abundance is 5x larger than that of standard, pre-JWST AGN of comparable bolometric luminosity (\citealt{shen20}; not shown in Fig.~\ref{fig_phi}), whereas it rapidly falls below it by a factor of $\sim7$ at $z\sim2.5$ (see Fig.~\ref{fig_phi}).

Despite this rapid decline as the Universe ages, the abundance of LRDs at cosmic noon remains significant, as also possibly confirmed by the results of \citet{ma25}, who leveraged ground-based photometric selection of LRDs over a $\sim3$ deg$^2$ area and derived optical luminosity functions at $z\sim2-3$. When correcting the observed 5500 $\AA$ luminosities in \citet{ma25} using an extinction of $A_{\rm V}\sim1.6$, i.e. the average value quoted by \citet{kokorev24} for their sample, and assuming a bolometric correction of 10 \citep{richards06}, their measurements convert into a space density of $z\sim2-3$ LRDs with $L_{\rm bol} \sim 5\times 10^{45}\ \ergs$ equal to $\sim 3 \times 10^{-6}$ Mpc$^{-3}$ dex$^{-1}$, which is very similar to our result.

This large abundance of LRDs at cosmic noon is particularly remarkable if compared with the expectations from recent models, which suggest that LRDs may probe the very first and rapid growth of BH seeds \citep{inayoshi25}. According to these models, BH seeds of mass $10^{4-5}$\Msun\ may grow via super-Eddington accretion in dense environments, and be able to reach masses of $10^{7-8}$\Msun\ in less than 10 Myr. After this early, rapid growth phase, because of the reduced mass supply from its close environment, the AGN transitions from super-Eddington to more typical accretion.

\citet{inayoshi25} demonstrated that the abundance of LRDs at $z>5$ and the abundance of AGN at $z<5$ can be reproduced by a model of stochastic BH accretion, in which the first one or two accretion episodes correspond to the LRD phase. In this model, the abundance of LRDs at $z\sim2.5$ is expected to fall short of that of standard AGN by three dex. This is in line with the expectations of other theoretical arguments that suggest that the formation rate of seed BHs is strongly suppressed at $z < 5$. This is expected to happen as metals diffuse throughout the intergalactic medium favoring cloud cooling and fragmentation, and, in turn, preventing direct collapse of large gas clouds into massive seed BHs \citep{bellovary11}.

However, at $z\sim2.5$, we measure the abundance of LRDs that is at most a factor of 7 lower than that of standard AGN. 
Our findings may then indicate that the formation of BH seeds does not halt at $z<5$, and remains somewhat efficient to at least cosmic noon.

\subsection{Active black hole mass function}
\label{sub:bhmassfunc}
We study the space density of LRDs at $z_{\rm mean} \sim 2.5$ as a function of the black hole mass. We use the black hole masses inferred from the \Ha\ line using the relation of \citet{reines&volonteri15}. \citet{wang25} adopt in their work the calibration of \citet{greene&ho05} to compute the \Ha - based black hole mass of RUBIES-BLAGN-1. We thus multiply their estimate by a factor of 2 to account for the systematic difference between the two relations.

The black hole masses of the three LRDs span over a mass bin of 1 dex centered at $M_{\rm BH, mean} \sim 6.3 \times 10^{8}\ \rm M_{\odot}$, and thus we have to divide the space density at $L_{\rm bol, mean}$ by a factor 2, as it was computed for a bin size of 0.5 dex ($\Phi(M_{\rm BH}) \sim 2 \times 10^{-6}\ \rm Mpc^{-3} dex^{-1}$). Besides the bin size, we do not expect that other effects impact the space density at $M_{\rm BH, mean}$. However, we note that, because of the color-magnitude selection box defined in Sect.~\ref{sub:spaceden}, and based on the rough correlation between F356W magnitude and BH mass observed in the few known LRDs at $z=2.0-3.5$ \citep{juodzbalis24,wang25,ma25,zhuang25},
we expect to sample high ($M_{\rm BH} \gtrsim 10^8\ \rm M_{\odot}$) black hole masses.

In Fig.~\ref{fig_phi} (right panel) we compare the space density of LRDs at $z_{\rm mean} \sim 2.5$ and $M_{\rm BH, mean} \sim 6.3 \times 10^{8}\ \rm M_{\odot}$ with the model black hole mass function (BHMF) of accreting BHs inferred from SDSS broad line QSOs at similar redshift ($z=2.65$; \citealt{shenkelly12}). For $M_{\rm BH, mean} \sim 6.3 \times 10^{8}\ \rm M_{\odot}$ and $z=2.65$ the SDSS data are indeed incomplete, therefore, we compare with the predicted\footnote{We note that the turnover at the low-mass end in Fig.~\ref{fig_phi} (right panel) may suffer from the poor sampling of low-mass BHs in the SDSS sample \citep{shenkelly12}} active BHMF of BLQSOs based on the observations. We see that our estimate is only a factor $\lesssim 2$ below the expectation for BLQSOs and consistent with it within the errors, thus implying that LRDs contribute in a way that is similar to BLQSOs to the population of black holes with a similar mass at $z\sim 2.5$.

We also compare our estimate with the active BHMF derived for JWST-discovered BLAGN at higher redshifts \citep{matthee24, taylor24}. We note that also in these works the black hole masses are computed using the calibration by \citet{reines&volonteri15}, so we do not apply any scaling factor to their active BHMFs. On the other hand, both \citet{matthee24} and \citet{taylor24} do not correct the \Ha\ luminosity for dust attenuation when estimating the black hole masses. Compared to the last bins of \citet{matthee24} and \citet{taylor24}, which partially overlap with our 1 dex-mass bin, we find that our estimate is a factor $\sim 3.8-4.4$ lower but consistent within the errors.

We note that \citet{matthee24} and \citet{taylor24} select their objects looking for broad-line emitters. Therefore, their active BHMFs include objects that are not LRDs. The fraction of LRDs is about $1/3$ of the sample from \citet{taylor24}. This means that their estimates would decrease when considering the BLAGN that are also LRDs. For the work by \citet{taylor24}, we evaluate a decrement of the active BHMF by a factor of less than $2$ in the last mass bin, due to the fact that the fraction of LRDs is higher than $1/3$ in that bin, which thus would made their estimates closer to ours. Overall, bearing in mind the different selections, this comparison points in the direction of a mild evolution only in the abundance of LRDs from high redshift to cosmic noon.

\section{Summary and conclusions}
\label{sec:concl}

In this work we report the study of BiRD, a LRD discovered at $z \sim 2.33$ in the J1030 field, using JWST/NIRCam data (both imaging and slitless spectroscopy). This source has interesting analogies with two other LRDs at cosmic noon, namely GN-28074 at $z = 2.26$ ('Rosetta Stone', \citealt{juodzbalis24}) and RUBIES-BLAGN-1 at $z = 3.1$ \citep{wang25}. The main conclusions of our work are:

\begin{itemize}
    
    \item Using the NIRCam/WFSS spectrum, we derive some physical properties of BiRD. The black hole mass, estimated with two independent methods starting from the broad \Pag\ emission, is of the order of $\sim 10^8\ \rm M_{\odot}$. We detected blueshifted \Hei\ absorpion ($\Delta v =-830_{-148}^{+131}$ \kms). A similar blueshifted feature is seen in the Rosetta Stone and in RUBIES-BLAGN-1. Depending on the absorber location, the \Hei\ column density is $N(\rm He\ I) \sim 0.5 - 1.2  \times 10^{14}\ \rm cm^{-2}$.\\

    \item We study the X-ray and radio properties of BiRD. The source is both radio and X-ray silent, with $L_{\rm 1.4\ GHz} < 3 \times 10^{39}\ \rm erg\ s^{-1}$ and $L_{2-10\ \rm keV, \rm rest} < 3.7 \times 10^{42}\ \rm erg\ s^{-1}$. Based on the bolometric luminosity $L_{\rm bol} \sim 3 \times 10^{45}\ \rm erg\ s^{-1}$, we find a lower limit to the X-ray bolometric correction $L_{\rm bol}/L_{2-10\ \rm keV, \rm rest} > 800$, consistent with the values of other JWST-discovered BLAGN.\\
    
    \item We estimate the space density of LRDs at cosmic noon, as a function of the bolometric luminosity $L_{\rm bol}$ and the black hole mass $M_{\rm BH}$. This is the first space density at that cosmic age that is constrained via spectroscopic data. We find $\Phi(L) = 4.0^{+4.0}_{-2.4} \times 10^{-6}$ Mpc$^{-3}$ dex$^{-1}$ at $L_{\rm bol,mean}= 5.4 \times 10^{45}\ \ergs$ and $z_{\rm mean}\sim2.5$. This value is only a factor $\sim 2-3$ lower than the value from UV-selected quasars at similar redshifts.\\
    
    \item Similarly, we find that for $M_{\rm BH}\sim 6.3 \times 10^{8}\ \rm M_{\odot}$, the space density $\Phi(M_{\rm BH}) \sim 2 \times 10^{-6}\ \rm Mpc^{-3} dex^{-1}$ is a factor $\sim 2$ only below the expectation for broad line QSOs in the same cosmic epoch. When we compare with JWST-discovered BLAGN at $z\sim 4.8$, our estimate is a factor $\sim 3.8 - 4.4$ below the active BHMF of LRDs with similar black hole masses, and consistent within the errors.
    
\end{itemize}

Overall, our results point to only a mild evolution in the abundance of LRDs with decreasing redshift, with their space density remaining significant at cosmic noon, comparable with that of BLQSOs.

In the future, we plan to further investigate the properties of BiRD using multiwavelength data. 
According to \citet{maiolino25}, the X-ray weakness of LRDs may be explained by dense, dust-poor or dust-free gas in BLR clouds. Future JWST/NIRSpec observations of the rest-frame optical spectrum of BiRD, with particular regard to the Balmer break and Balmer absorptions, may test whether this scenario also applies in this system.
Regarding the dust properties, follow-up observations of BiRD in the mid-IR and millimeter regimes with JWST/MIRI and ALMA, respectively, are essential to place robust constraints on the amounts of hot and cold dust in this system. MIRI observations will also allow us to detect the \Hei\ $+$ \Pag\ lines in other LRDs at higher redshifts, to verify if \Hei\ absorptions are widespread in BLAGN also before the cosmic noon.
Another interesting development of our study would be to extend the search for objects similar to BiRD using larger areas of the sky. This will enable the sampling of a wider luminosity/black hole mass space for the computation of the luminosity function/BHMF of LRDs at cosmic noon. All these future expansions of this work are pivotal if we want to shed more light on the elusive nature of LRDs.

\begin{acknowledgements} 
We thank the referee for their useful and constructive comments.
We thank Guido Risaliti for useful discussion and for kindly providing us with the data-points of Fig. 6. We acknowledge support from the INAF 2022/2023 "Ricerca Fondamentale" grants. 
K.I. acknowledges support under the grant PID2022-136827NB-C44 provided by MCIN/AEI/10.13039/501100011033 / FEDER, UE.
I.J. and R.M. acknowledges support by the Science and Technology Facilities Council (STFC), by the ERC through Advanced Grant 695671 “QUENCH”, and by the UKRI Frontier Research grant RISEandFALL.
I.J. acknowledges support also by the Huo Family Foundation through a P.C. Ho PhD Studentship. 
R.M. also acknowledges funding
from a research professorship from the Royal Society. M.S. acknowledges financial support from the Italian Ministry for University and Research, through the grant PNRR-M4C2- I1.1-PRIN 2022-PE9-SEAWIND: Super-Eddington Accretion: Wind, INflow and Disk-F53D23001250006-NextGenerationEU.
This work is based on observations made with the NASA/ESA/CSA James Webb Space Telescope. The data were obtained from the Mikulski Archive for Space Telescopes at the Space Telescope Science Institute, which is operated by the Association of Universities for Research in Astronomy, Inc., under NASA contract NAS 5-03127 for JWST. These observations are associated with the GTO program 1243.
\end{acknowledgements}

\bibliographystyle{aa}
\bibliography{main}
\end{document}